\documentclass[10pt]{article}

\usepackage[T1,T2A]{fontenc}
\usepackage[cp1251]{inputenc}
\usepackage{textcomp}
\usepackage[russian,english]{babel}
\usepackage[centertags]{amsmath}
\usepackage[mediummath]{nccmath}
\usepackage{amsfonts}
\usepackage{amssymb}
\usepackage[driverfallback=dvipdfm]{hyperref}
\usepackage{graphicx}
\usepackage[numbers,sort&compress]{natbib}

\usepackage{indentfirst}
\usepackage{amsmath, amssymb, graphics, setspace}
\usepackage{caption} 

\usepackage{xcolor} 
\newcommand{\paperinitialization}[6]{%
	\setlength{\hoffset}{-1in} \setlength{\voffset}{-1in}
	\setlength{\headheight}{0pt} \setlength{\headsep}{0pt}
	\setlength{\marginparsep}{#5}
	\setlength{\marginparwidth}{#4}\addtolength{\marginparwidth}{-#6}
	\setlength{\topmargin}{#1}
	\setlength{\oddsidemargin}{#3} \setlength{\evensidemargin}{#4}
	\setlength{\textwidth}{\paperwidth}%
	\addtolength{\textwidth}{-#3}
	\addtolength{\textwidth}{-#4}%
	\setlength{\textheight}{\paperheight}%
	\addtolength{\textheight}{-#1}
	\addtolength{\textheight}{-#2}%
}%

\paperinitialization{20mm}{20mm}{20mm}{15mm}{2pt}{10pt}

\DeclareMathOperator{\sgn}{sgn}

\newcommand{\bs}{\boldsymbol}

\newcommand{\sixj}[6]{%
	\begin{Bmatrix}
		#1 & #2 & #3 \\
		#4 & #5 & #6
\end{Bmatrix}}

\newcommand{\e}{\varepsilon}
\newcommand{\vf}{\varphi}

\newcommand{\s}{\sigma}

\newcommand{\al}{\alpha}

\newcommand{\ga}{\gamma}

\newcommand{\de}{\delta}
\newcommand{\De}{\Delta}

\newcommand{\la}{\lambda}
\newcommand{\La}{\Lambda}

\newcommand{\spx}{\mathbf{x}}

\newcommand{\spk}{\mathbf{k}}
\newcommand{\spe}{\mathbf{e}}

\newcommand{\spP}{\mathbf{P}}
\newcommand{\spr}{\mathbf{r}}

\begin{document}

\allowdisplaybreaks[4]
\setlength{\unitlength}{1pt}

\title{{\Large\textbf{Multipolar transitions excited by twisted photons in heavy quarkonia}}}
\date{}	

\author{P.S. Korolev\thanks{E-mail: \texttt{kizorph.d@gmail.com}},\; V.A. Ryakin\thanks{E-mail: \texttt{vlad.r.a.phys@yandex.ru}}\\[0.5em]
	{\normalsize  Physics Faculty, Tomsk State University, Tomsk 634050, Russia}\\[0.5em]
	{\normalsize Mathematics and Mathematical Physics Division,}\\ {\normalsize Tomsk Polytechnic University, Tomsk 634050, Russia}
}

\maketitle

\begin{abstract} 
	 In this study, we investigate the excitation of multipole transitions in quarkonium systems by high-energy photons carrying orbital angular momentum. We derive explicit expressions for the amplitude and probability of photoexcitation incorporating the dynamics of the quarkonium center of mass. It is shown that the leading contribution to the transitions induced by twisted photons comes from the multipole $j=|m_\gamma|$ and $m_\gamma=l'-l$ in the long-wave approximation. Despite the fact that similar transitions can be excited by plane-wave photons, twisted photons with $|m_\gamma|>1$ enable the isolation of higher multipoles offering a unique spectroscopic tool. The fulfillment of the derived selection rules imposes constraints on the parameters of the wave packet of the quarkonium center of mass, since, in general, the angular momentum can be transferred to the center of mass. As a case study, we examine the octupole transition $1^{3}G_3 \rightarrow 3^{3}P_0$ in the charmonium system. We also compare the probabilities of excitation of this multipole transition by plane-wave and twisted photons with different polarizations.

\end{abstract}

\section{Introduction}

The structure and the energy spectrum of heavy quarkonia, the bound states of a heavy quark and an antiquark, provide an essential testing ground for quantum chromodynamics (QCD) in the non-perturbative regime. In 1980, the first spin-singlet state $\eta_c$ in the $M1$ suppressed transition $\psi(2S)\rightarrow \eta_c \gamma$ \cite{Partridge1980} was detected.  Then in 2002, the Belle experiment observed the $\eta_c(2S)$ state in both $B$-meson decays \cite{Choi2002} and in the exclusive process $e^+ + e^- \to J/\psi + \eta_c(2S)$ \cite{Abe2002}. For the theoretical description of the observed quarkonia properties various potential models inspired by quantum electrodynamics were proposed \cite{Eichten1978, quigg1979, Richardson1979}. Despite the success of such models in describing the lowest excited states of quarkonia, many discrepancies between theoretical predictions and experimental results have been discovered in recent decades \cite{brambilla2020, eichten2019, gross2023}, especially for highly excited states of charmonium and bottomonium. These discrepancies point to the presence of as yet unexplored internal dynamics.

Electromagnetic transitions of higher multipolarities can become a powerful tool for studying the internal structure of heavy quarkonia. Such transitions carry additional information about spatial distribution and spin-dependent effects in the system. However, such transitions are strongly suppressed when induced by the plane waves and, as a rule, they cannot be distinguished in the presence of the dominant dipole transitions. This limits our ability to accurately study the spectral characteristics. In this paper, we study the possibility of inducing and selectively exciting higher-multipole transitions in heavy quarkonia using twisted photons, quantum states of light that carry orbital angular momentum (OAM). While twisted photons have been used in atomic \cite{grinter2008,matula2013,afanasev2013,afanasev2016,Solyanik-Gorgone2019,shulz2019,shulz2020,peshkov2023, KKR2025} and nuclear physics \cite{KS2024, LuZhi2023}, their application to quarkonia system remains unexplored. In the paper \cite{KR2023}, the twisted photons have been employed for excitation of Wannier excitons -- a planar hydrogen-like binary system -- providing a highly sensitive tool for testing the electron-hole interaction. Here, we demonstrate how they can selectively excite higher multipole transitions in heavy quarkonia.

%

The phenomenological approach to the study of the heavy quarkonium system is based on the use of analogy with the hydrogen atom or positronium. For this purpose, the Schr\"{o}dinger equation for the quark-antiquark pair is solved with the scalar potential $V(r)$ depending only on the distance $r$ between the quark and the antiquark. The use of such a model for heavy quarkonia was quite successful because this system is approximately nonrelativistic. The nonrelativistic approach is justified by the fact that the mass of the $b$-quark (and to a lesser extent of the $c$-quark) is much larger than $\Lambda_{QCD}$.

To date, many models of the quark-antiquark pair interaction potential have been proposed \cite{Eichten1975, Eichten1976, Lane1976, Eichten1978, Eichten1980, quigg1979, Richardson1979, Buchmuller1981, barchielli1990, eiglsperger2007, Brambilla2001, Brambilla2005, Mutuk2018}. The most widely-used one is the Cornell potential \cite{Eichten1975, Eichten1976, Lane1976, Eichten1978, Eichten1980}. It consists of two terms: the Coulomb-type term that can be obtained by means of perturbative Quantum Chromodynamics (QCD) by considering one gluon exchange diagrams for quark-antiquark scattering \cite{Gromes1991} and a linear with respect to $r$ term responsible for confinement. Other popular models of the quark interaction potential include the Rosner potential, the harmonic potential, the logarithmic potential, the step potential, and all their various linear combinations (for a review, see \cite{quigg1979}). When relativistic corrections are taken into account, the standard contributions depending on the spin and relative orbital momentum of quarks are added to the seed potential; they are divided into the spin-orbit interaction, the spin-spin interaction, and the tensor interaction. These contributions follow from the Breit equation \cite{Breit1929, Breit1932}. 

Excited states of quarkonium can decay to the ground state with emission of a photon. Since the size of quarkonium is significantly smaller than the wavelength of the emitted photon the electric dipole transitions, $E1$, are the most probably realized. The second largest contribution to the radiation probability is the magneto-dipole transition $M1$, which is caused by the presence of the magnetic moment of the quark - antiquark system. As a rule, only these transitions are taken into account for description of processes of photon radiation by quarkonia. In the present paper we are primarily interested in the transitions $Ej$ and $Mj$ of higher multipolarity in the reverse process of photoexcitation.

In the recent years the twisted photons have drawn a lot of attention due to their unique properties stemming from the nonzero orbital angular momentum (see for review \cite{PadgOAM25}). Such photons represent a promising tool for parallel signal coding in (quantum) telecommunication \cite{Roadmap16,New18,OAMPM}. They can also be used in optical tweezers and in controlling the rotational degrees of freedom of nanoparticles, molecules, atoms, and nuclei \cite{Roadmap16,SerboNew,New19}. In the present paper, we focus on the other application of twisted photons, namely, we consider them as a tool for photoexcitation of higher multipole transitions in quarkonia that are forbidden in the dipole approximation. Energies of such transitions in quarkonia are usually of the order of 100 MeV. The twisted photons with such energies can be created by using the inverse Compton scattering \cite{Jentshura2011_1, Jentshura2011_2, BKL2019, Ivanov2022, Guo2023} or channeling radiation \cite{ABKT2018,BKT2022}.


The paper is organized as follows. In Section \ref{SectionMainFormular}, we give the general formulas describing the quarkonium system and transitions in them. Section \ref{SectionProbability} is devoted to evaluation of  the probability of photoexcitation of quarkonium by a twisted photon. In Section \ref{SectionNumerical}, we apply the derived formulas to the octupole transition in the charmonium system. In conclusion section, we summarize the results. In Appendices  \ref{SmallKappaEvaluation} and \ref{BigKappaEvaluation} we calculate the contributions to the matrix elements of the multipole transitions coming from the relative motion of the quarks and from the quarkonium center of mass. Throughout the text we use the system of units such that $\hbar=c=1$ and $e^2=4\pi\al$, where $\hbar$ is the Planck constant, $c$ is the speed of light, $e$ is the electron charge, and $\al$ is the fine structure constant. We also use the notations for the $1$, $2$, $3$ axes and $x$, $y$, $z$ interchangeably.

\section{Main formulas}\label{SectionMainFormular}

Let us consider a non-relativistic quark-antiquark system interacting with an electromagnetic field
\begin{equation}\label{H_initial}
	H=\frac{(\mathbf{p}_1-q \mathbf{A}(\mathbf{r}_1))^2}{2m_1}+\frac{(\mathbf{p}_2-\bar{q} \mathbf{A}(\mathbf{r}_2))^2}{2m_2}+V_0(|\mathbf{r}_1-\mathbf{r}_2|) + H_\text{EM},
\end{equation}
where $q$, $\bar{q}=-q$ are the charges of quark and antiquark, respectively, $H_{EM}$ is the Hamiltonian of a free electromagnetic field. In the general derivation we keep $m_1$ and $m_2$ arbitrary, which allow the formalism to describe both equal-mass systems such as charmonium ($c\bar{c}$) and bottomonium ($b\bar{b}$), as well as unequal-mass systems such as $B_c$ mesons. In the next section we set $m_1=m_2$. The free quantum electromagnetic field operator in vacuum in the Coulomb gauge reads as
\begin{equation}
	A_i(\mathbf{r},t)=\sum_{\ga}\hat{c}_\ga \Psi_{\ga i}(\mathbf{r})e^{-ik_{0\ga}t}+\sum_{\ga}\hat{c}^\dagger_\ga \Psi^*_{\alpha i}(\mathbf{r})e^{ik_{0\ga}t},\quad i=1,2,3,
\end{equation} 
where $\hat{c}^{\dagger}_\ga,$ $\hat{c}_\ga,$ are creation and annihilation operators satisfying the standard commutation relations
\begin{equation}
	[\hat{c}_\ga,\hat{c}_\beta ] = [\hat{c}^{\dagger}_\ga,\hat{c}^{\dagger}_\beta ] = 0	,\qquad  [\hat{c}_\ga,\hat{c}^{\dagger}_\beta ]\ = \delta_{\ga,\beta}.
\end{equation}
The plane-wave mode functions are written as
\begin{equation}
	\mathbf{\Psi}_{\ga}(\mathbf{r}) = \frac{\mathbf{f}_\lambda(\mathbf{k})}{\sqrt{2k_0 V}} e^{i\lambda\sgn(k_3)\phi_k} e^{i\mathbf{k}\mathbf{r}},\quad \ga = (\lambda,\mathbf{k}),\quad \sum_{\ga} = \sum_{\la =\pm 1}\int \frac{Vd\spk}{(2\pi)^3},
\end{equation}
where the circular polarization vector has the form
\begin{equation}\label{VecPol}
	\mathbf{f}_\lambda(\mathbf{k})=(\cos \phi_{k} \cos \theta_{k}-i\lambda \sin \phi_{k},\sin\phi_{k}\cos\theta_{k}+i\lambda\cos\phi_{k},-\sin\theta_{k})/\sqrt{2},
\end{equation}
and $k_\perp = |k_1+i k_2|$, $\phi_{k}=\arg(k_1+ik_2)$, $\sin\theta_{k}:=k_\perp/k_0$, $\cos\theta_{k}=k_3/k_0$. Note that $ \mathbf{f}_\lambda(\mathbf{k}) $, generally speaking, does not transform as a vector under rotations, moreover, it is not a smooth function in the limit $k_\perp \rightarrow 0$. In order to eliminate this problem, the phase factor $e^{i\lambda\sgn(k_3)}$ is introduced. Since the general phase does not affect the probabilities of processes involving plane waves, this factor is usually omitted. However, in the case of non-trivial wave packets, it is necessary to take into account all phase factors.

Expanding the brackets in \eqref{H_initial}, we arrive at the Hamiltonian
\begin{equation}\label{H_main}
	H=H_0+H_\text{int}^{(1)}+H_\text{int}^{(2)},
\end{equation}
where
\begin{equation}
	\begin{split}
		H_0 &= \frac{\mathbf{p}_1^2}{2m_1}+\frac{\mathbf{p}_2^2}{2m_2}+V_0(|\mathbf{r}_1-\mathbf{r}_2|)+H_\text{EM},\\
		H_\text{int}^{(1)} &= -\frac{q}{2m_1}\big(\mathbf{A}(\mathbf{r}_1)\mathbf{p}_1+\mathbf{p}_1\mathbf{A}(\mathbf{r}_1) \big)+
		\frac{q}{2m_2}\big(\mathbf{A}(\mathbf{r}_2)\mathbf{p}_2+\mathbf{p}_2\mathbf{A}(\mathbf{r}_2) \big),\\
		H_\text{int}^{(2)} &= \frac{q^2}{2m_1}\mathbf{A}^2(\mathbf{r}_1)+\frac{q^2}{2m_2}\mathbf{A}^2(\mathbf{r}_2).
	\end{split}
\end{equation}
The last term in the Hamiltonian \eqref{H_main}, which is proportional to  $A^2$, is of higher order with respect to the coupling constant than the first two terms and is related to the two photon processes. Therefore, it can be discarded when considering light emission and absorption in the leading order of perturbation theory. Taking into account the divergence-free nature of the electromagnetic field, we can write
\begin{equation}
	H_\text{int}^{(1)}=-\frac{q}{m_1}\mathbf{A}(\mathbf{r}_1)\mathbf{p}_1 +
	\frac{q}{m_2}\mathbf{A}(\mathbf{r}_2)\mathbf{p}_2.
\end{equation}

It is convenient to switch to the center-of-mass coordinates
\begin{equation}
	\mathbf{P}=\mathbf{p}_1+\mathbf{p}_2, \quad \mathbf{R}=\frac{m_1\mathbf{r}_1+m_2\mathbf{r}_2}{m_1+m_2} ,\quad M_q=m_1+m_2,
\end{equation}
and
\begin{equation}
	\mathbf{p}=\frac{m_2\mathbf{p}_1-m_1\mathbf{p}_2}{m_1+m_2},\quad \mathbf{r}=\mathbf{r}_1-\mathbf{r}_2,\quad \mu = \frac{m_1 m_2}{m_1+m_2}.
\end{equation}
Hence the Hamiltonian takes the form
\begin{equation}
	\begin{split}
		H_{0} &=H_{q\bar{q}}+H_{CM} + H_\text{EM},\quad H_{q\bar{q}}=\frac{\mathbf{p}^2}{2\mu}+V_0(|\mathbf{r}|),\quad H_{CM}=\frac{\mathbf{P}^2}{2M_q},\\
		H_\text{int}^{(1)} &=q\big(\mathbf{A}(\mathbf{R}-\frac{m_1}{M_q}\mathbf{r})-\mathbf{A}(\mathbf{R}+\frac{m_2}{M_q}\mathbf{r})\big) \frac{\mathbf{P}}{M_q}-q\big( \frac{1}{m_2}\mathbf{A}(\mathbf{R}-\frac{m_1}{M_q}\mathbf{r}) + \frac{1}{m_1}\mathbf{A}(\mathbf{R}+\frac{m_2}{M_q}\mathbf{r})  \big)\mathbf{p}.
	\end{split}
\end{equation}

We represent the eigenfunctions for the Hamiltonian $ H_0 $ in the form $ \Psi_\spP(\mathbf{R}) \psi_{nlm}(\mathbf{r})$. The functions $ \Psi_\spP(\mathbf{R}) $ and $ \psi_{nlm}(\mathbf{r}) $ satisfy the following equations
\begin{equation}\label{totShr}
\begin{split}
	\bigg(-\frac{\Delta_{\mathbf{R}}}{2M_q}-E_{CM}\bigg)\Psi_\spP(\mathbf{R})=0,\\
	\bigg(-\frac{\Delta_{\mathbf{r}}}{2\mu}+V_0(|\mathbf{r}|)-E_{nlm}\bigg)\psi_{nlm}(\mathbf{r})=0,
\end{split}
\end{equation}
where $ E_{CM} := \frac{\mathbf{P}^2}{2M_q} $. We write the solution of the first equation as
\begin{equation}
	|\spP\rangle := \Psi_\spP(\mathbf{R}) = \frac{1}{\sqrt{V}}e^{i\mathbf{P}\mathbf{R}}.
\end{equation}
The solution of the second equation in \eqref{totShr} is presented in the standard form
\begin{equation}
	|nlm\rangle := \psi_{nlm}(r,\theta,\phi) = R_{nl}(r)Y_{lm}(\theta,\varphi),
\end{equation}
where $Y_{lm}(\theta,\varphi)$ are the spherical harmonics, and $l=0,1,2,\dots$, $m=0,\pm 1,\dots, \pm l$. The radial wave function $R_{nl}(r)$ satisfies the equation
\begin{equation}\label{RadShr}
	\bigg[-\frac{1}{2\mu} \bigg(\frac{\partial^2}{\partial r^2} + \frac{2}{r}\frac{\partial}{\partial r}\bigg) + \frac{l(l+1)}{2\mu r^2} + V_0(r)\bigg] R_{nl}(r)= E_{nl}R_{nl}(r).
\end{equation}  

There are various models that approximately describe the quarkonium spectrum, including those that take into account relativistic corrections. The most successful model among the simple ones is described by the Cornell potential \cite{Eichten1975, Eichten1976, Lane1976, Eichten1978, Eichten1980}
\begin{equation}\label{CornellPotential}
	V_0(r)=-\frac{4\alpha_s}{3r}+\sigma r.
\end{equation} 
Here $\alpha_s$ is a running coupling constant and $\sigma$ is a string tension. The first term in this potential is called the Coulomb part and it can be derived explicitly by considering the one-gluon exchange diagram between a quark and an antiquark \cite{Gromes1991}. The second term is responsible for the confinement of quarks. Note that such a potential does not depend on spin and cannot fully reproduce the observed spectra of quarkoniums. However it describes with good accuracy transitions between the states with the same spin quantum number. The Hamiltonian \eqref{H_initial} is nonrelativistic and we will take relativistic corrections into account by using perturbation theory. To this end, we introduce the Breit potential \cite{Breit1929, Breit1932}, which was originally developed to describe the energy spectrum of positronium states accounting for spin dependent corrections. The Breit potential, when adapted to the quarkonium system, shows excellent agreement with the observed energy spectrum of quarkonia for the lowest energy levels, see \cite{eiglsperger2007}. For the $q\bar{q}$ system, it has the following form
\begin{equation}\label{BreitPotential}
	V^{\text{Breit}}(\mathbf{r},\mathbf{p})= \frac{\pi\alpha_s}{3\mu^2}\de(\mathbf{r})-\frac{\alpha_s}{6\mu^2}\big[ \frac{\mathbf{p}^2}{r}+\frac{(\mathbf{r}\mathbf{p})^2}{r^3}\big] + \frac{\alpha_s}{3\mu}\big[\frac{8\pi}{3}\delta(\mathbf{r}) (\mathbf{s}_1\mathbf{s}_2) + \frac{3 (\mathbf{s}_1\mathbf{r}) (\mathbf{s}_2\mathbf{r})}{r^5}-\frac{(\mathbf{s}_1\mathbf{s}_2)}{r^3}\big] + \frac{\alpha_s}{2\mu^2}\frac{(\mathbf{r}\times \mathbf{p})(\mathbf{s}_1+\mathbf{s}_2)}{r^3},
\end{equation}
where $\mathbf{s}_1$, $\mathbf{s}_2$ are the quark spin operators. Therefore the energy corrections are calculated as
\begin{equation}
	\delta E_{nlsj} = \langle nlm| V^{\text{Breit}}(\mathbf{r},\mathbf{p})|nlm\rangle,
\end{equation}
and the corrections to the wave function are given by
\begin{equation}
	|nlmsj\rangle :=\psi_{nlmsj}(r,\theta,\phi) = (H_{q\bar{q}}-E_{nl})^{-1}\big[\delta E_{nlsj}-V^{\text{Breit}}(\mathbf{r},\mathbf{p})\big]|nlm\rangle,
\end{equation}
where $j,s$ are the eigenvalues of the total angular momentum operator $\mathbf{J}$ and the total spin operator $\mathbf{S}$, respectively. We will use the expressions for the corrections in the numerical simulation section.

Suppose that the system is prepared in the state $|\text{in}\rangle$ at the time $t_1$, and it is measured in the state $|\text{out}\rangle$ at time $t_2>t_1$ 
\begin{equation}\label{inoutStates}
	\begin{split}
		&|\text{in}\rangle = \sum_{\ga}\chi_\ga e^{-ik_{0\ga}t_1}\hat{a}^{\dagger}_\ga|0\rangle\int \frac{V^{1/2} d\mathbf{P}}{(2\pi)^{3/2}} e^{-iE_\text{in}t_1} g(\mathbf{P})|\mathbf{P}\rangle |nlm\rangle ,\\
		&|\text{out}\rangle =e^{-iE_\text{out}t_2}|0\rangle|\mathbf{P'}\rangle |n'l'm'\rangle,
	\end{split}
\end{equation}
where $ \chi_\ga $ and $ g(\spP) $ define the wave packet profiles of the photon and of the center of mass of the quarkonium, respectively. They obey the normalization conditions
\begin{equation}\label{norm_conds}
	\sum_\ga|\chi_\ga|^2=1,\qquad\int d\spP |g(\spP)|^2=1.
\end{equation}
Furthermore,
\begin{equation}
	\begin{split}
		(\hat{H}_{q\bar{q}} +\hat{H}_{CM})|\mathbf{P}\rangle |nlm\rangle=E_{\text{in}}|\mathbf{P}\rangle |nlm\rangle &, \quad (\hat{H}_{q\bar{q}} +\hat{H}_{CM})|\mathbf{P'}\rangle |n'l'm'\rangle =E_{\text{out}}|\mathbf{P'}\rangle |n'l'm'\rangle,
		\\
		E_{\text{in}} := E_{nlm} +\frac{\mathbf{P}^{2}}{2 M_q}&,\quad E_{\text{out}} := E_{n'l'm'} +\frac{\mathbf{P'}^{2}}{2 M_q}.
	\end{split}
\end{equation}
Then the transition amplitude from the state $|\text{in}\rangle$ to the state $|\text{out}\rangle$,
\begin{equation}
\langle\text{out}|\hat{U}_{t_2,t_1}|\text{in}\rangle,
\end{equation}
is determined by the evolution operator
\begin{equation}
\hat{U}_{t_2,t_1}=\hat{U}^0_{t_2,0} \hat{S}_{t_2,t_1} \hat{U}^0_{0,t_1},\quad \hat{S}_{t_2,t_1}=\text{Texp}\bigg\{-i\int_{t_1}^{t_2}dt \hat{H}_{\text{int}}(t)\bigg\},
\end{equation}
where $\hat{U}^0_{t_2,t_1}$ is the free evolution operator
\begin{equation}
\hat{U}^0_{t_2,t_1}= e^{-i \hat{H}_0(t_2-t_1)},
\end{equation}
and $\hat{H}_{\text{int}}(t)$ is the interaction Hamiltonian in interaction representation
\begin{equation}
\hat{H}_{\text{int}}(t) = e^{i\hat{H}_0 t} \hat{H}_\text{int} e^{-i\hat{H}_0 t}.
\end{equation}

In the first Born approximation we have
\begin{equation}
\hat{S}_{t_2,t_1}\approx 1-i\int_{t_1}^{t_2}dt \hat{H}_{\text{int}}(t).
\end{equation}
Then the leading non-trivial contribution to the transition amplitude takes the form
\begin{equation}\label{general_ampl}
\begin{split}
	-i\langle\text{out}| \hat{U}^0_{t_2,0}  \int_{t_1}^{t_2}dt e^{i\hat{H}_0 t}	H_\text{int}^{(1)} e^{-i\hat{H}_0 t} \hat{U}^0_{0,t_1} |\text{in}\rangle =-  &iq \sum_{\ga  }\int_{t_1}^{t_2}dte^{i(E_\text{out}-E_\text{in}-k_{0\ga  })t}(K_\ga  ^{-}+\kappa_\ga ^{-})-\\
	-&iq \sum_{\ga  }\int_{t_1}^{t_2}dte^{i(E_\text{out}-E_\text{in}+k_{0\ga  })t}(K_\ga  ^{+}+\kappa_\ga  ^{+}),
\end{split}
\end{equation}
where
\begin{equation}
\begin{split}
	K_\ga  ^{+}&= M_q^{-1}\langle\text{out}| \hat{c}_\ga  ^\dag \big(\Psi_{\ga   i}^{*}(\mathbf{R}-\frac{m_1}{M_q}\mathbf{r}) -\Psi_{\ga   i}^{*}(\mathbf{R}+\frac{m_2}{M_q}\mathbf{r}) \big)P_i |\text{in}\rangle ,
	\\
	K_\ga  ^{-}&= M_q^{-1}\langle\text{out}| \hat{c}_\ga   \big(\Psi_{\ga   i}(\mathbf{R}-\frac{m_1}{M_q}\mathbf{r}) -\Psi_{\ga   i}(\mathbf{R}+\frac{m_2}{M_q}\mathbf{r}) \big)P_i |\text{in}\rangle,
	\\
	\kappa_\ga  ^{+}&=- \langle\text{out}| \hat{c}_\ga  ^\dag \big(m_2^{-1}\Psi_{\ga   i}^{*}(\mathbf{R}-\frac{m_1}{M_q}\mathbf{r}) +m_1^{-1}\Psi_{\ga   i}^{*}(\mathbf{R}+\frac{m_2}{M_q}\mathbf{r}) \big)p_i |\text{in}\rangle ,
	\\
	\kappa_\ga  ^{-}&=- \langle\text{out}| \hat{c}_\ga   \big(m_2^{-1}\Psi_{\ga   i}(\mathbf{R}-\frac{m_1}{M_q}\mathbf{r}) +m_1^{-1}\Psi_{\ga   i}(\mathbf{R}+\frac{m_2}{M_q}\mathbf{r}) \big)p_i |\text{in}\rangle.
\end{split}
\end{equation}
The phases of the initial and final states \eqref{inoutStates} are selected in such a way that they cancel out the phases arising from the action of the free evolution operator. The matrix elements $K_\ga ^{-}$ and $\kappa_\ga ^{-}$ are associated with the process of absorption of one photon with energy $k_{0\ga }$, and $K_\ga ^{+}$ and $\kappa_\ga ^{+}$ correspond to the inverse process -- spontaneous emission of one photon. We will further assume that $ m_1 = m_2$.

Finally, the probability of photoabsorption of a photon by quarkonium in the non-relativistic approximation has the form
\begin{equation}\label{Prob_main}
	\begin{split}
		P_{n'l'm',nlm} = \int \frac{V d\mathbf{P'}}{(2\pi)^3}|\langle\text{out}|\hat{U}_{\infty,-\infty}|\text{in}\rangle|^2= q^2 \int \frac{Vd\spP'}{(2\pi)^3} \big|\sum_{\ga} 2\pi\de(E_\text{out}-E_\text{in}-k_{0\ga}) (K_\ga^{-}+\kappa_\ga^{-})\big|^2 ,
	\end{split}
\end{equation}
where it is also assumed that the momentum of the center of mass $\spP'$ is not measured and we integrate the probability with respect to it.

\section{Probability of photoexcitation}\label{SectionProbability}

Let us consider the amplitude of absorption of one photon by quarkonium. The calculations of matrix elements $\kappa_\ga^{-}$ and $K_\ga^{-}$ in the multipole expansion are given in appendices \ref{SmallKappaEvaluation} and \ref{BigKappaEvaluation}, respectively. Thus we have
\begin{equation}
	\begin{split}
		&\kappa_\ga^{-} =  \frac{\chi_\ga e^{i\lambda \phi_k} }{\sqrt{2 k_{0\ga}}}\int\frac{ d\mathbf{P}}{(2\pi)^{3/2}} g(\mathbf{P}) \langle \mathbf{P}'|e^{i\mathbf{R}\mathbf{k}_\ga}|\mathbf{P}\rangle
		4\pi \sum_{J\Lambda M}i^{J+\Lambda}\bigg(\mathbf{f}_\lambda(\mathbf{k}_\ga) \mathbf{Y}_{JM}^{(\Lambda)*}(\theta_{k},\phi_{k})\bigg) C_{lm JM}^{l'm'}  B_{J}^{(\Lambda)}, \\
		&K_\ga^{-}=   \frac{\chi_\ga e^{i\lambda\phi_k} }{\sqrt{2 k_{0\ga}}}\int\frac{ d\mathbf{P}}{(2\pi)^{3/2}} g(\mathbf{P})\langle \mathbf{P}'|e^{i\mathbf{R}\mathbf{k}_\ga}|\mathbf{P}\rangle 4\pi \sum_{J\Lambda M}i^{J+\Lambda} \bigg(\mathbf{f}_\lambda(\mathbf{k}_\ga) \mathbf{Y}_{JM}^{(\Lambda)*}(\theta_{k},\phi_{k})\bigg)	\bigg(\mathbf{W}^{(\Lambda)}_{JM}\mathbf{P}\bigg),
	\end{split}
\end{equation}
where $J=1,\cdots$,$\infty$, $M=0,\pm1,\cdots, \pm J$, $\Lambda=1$ for electric ($E$) transitions, $\Lambda=0$ for magnetic ($M$) transitions, and expressions for the $ B_{J}^{(\Lambda)}$ and $ \mathbf{W}^{(\Lambda)}_{JM}$ are presented in \eqref{B_JM^La} and \eqref{W_JM}, respectively. Notice that the selection rules for the photoexcitation process are contained in these objects. In particular, the matrix element $\kappa_\ga^{-}$ is not zero when 
\begin{equation}\label{BSelRul}
\begin{split}
&M = m' - m,\qquad J \geq |m'-m|,\qquad |l' - l| \leq J \leq l' + l, \\
&  l' - l + J + \La\; \text{odd},\quad J + \La \; \text{even},\\
\end{split}
\end{equation}
whereas the matrix element $K_\ga^{-}$ does not vanish when 
\begin{equation}\label{WSelRul}
\begin{split}
&M = m' - m + \s,\qquad J \geq |m'-m + \s|, \qquad |l' - l| \leq J \pm \La \leq l' + l,\\
&l' - l + J + \La \; \text{even}, \qquad J + \La \; \text{odd},
\end{split}
\end{equation}
where $ \s $ takes values from the set $\{-1,0,1\} $. As can be seen, the selection rules imply that only transitions with odd $|l'-l|$ are allowed. In particular, these selection rules forbid the transition $1^1 S_0 (\eta_c(1S))\rightarrow 1^3 S_1 (J/\Psi)$, the inverse to which is actually observed in the experiment. For this transition, the relativistic corrections make a significant contribution to the amplitude and modify the above selection rules. This transition, however, is dipole and is not of interest within the scope of this paper. As will be discussed later, we will be interested in low-energy multipole transitions obeying the selection rules \eqref{BSelRul} with the same spin quantum number in the initial and the final state.

The transition amplitude reads
\begin{equation}
	\begin{split}
		\langle\text{out}|\hat{U}_{\infty,-\infty}|\text{in}\rangle=&-\frac{2 iq (2\pi)^4}{V\sqrt{2\pi}} \sum_\gamma \frac{\chi_\ga e^{i\lambda\phi_k}}{\sqrt{2 k_{0\ga}}} \int d\mathbf{P}g(\mathbf{P})\delta(\e +\frac{\mathbf{P}'^2}{2M_q}-\frac{\mathbf{P}^2}{2M_q}-k_{0\gamma})\delta(\mathbf{P}'-\mathbf{P}-\mathbf{k}_\gamma)\times
		\\
		&\times \sum_{J\Lambda M}i^{J+\Lambda} \bigg(\mathbf{f}_\lambda(\mathbf{k}) \mathbf{Y}_{JM}^{(\Lambda)*}(\theta_{k},\phi_{k})\bigg)	\bigg(\mathbf{W}^{(\Lambda)}_{JM}\mathbf{P}+ C_{lm JM}^{l'm'} B_{J}^{(\Lambda)}\bigg), 
	\end{split}
\end{equation}
where $\e:=E_{n'l'm'}-E_{nlm}$ is the photoexcitation energy. 

On performing integral with respect to $\mathbf{P}$ by virtue of the $ \de $-function, we have
\begin{equation}
\begin{split}
	\langle\text{out}|\hat{U}_{\infty,-\infty}|\text{in}\rangle=-\frac{2 iq (2\pi)^4}{V\sqrt{2\pi}} \sum_\gamma \frac{\chi_\ga e^{i\lambda\phi_k}}{\sqrt{2 k_{0\ga}}} g(\mathbf{P}'-\mathbf{k}_\gamma)\delta(\e+\frac{\mathbf{P}'\mathbf{k}_\gamma}{M_q}-\frac{\mathbf{k}_\gamma^2}{2M_q} -k_{0\gamma})\times
	\\
	\times \sum_{J\Lambda M}i^{J+\Lambda} \bigg(\mathbf{f}_\lambda(\mathbf{k}) \mathbf{Y}_{JM}^{(\Lambda)*}(\theta_{k},\phi_{k})\bigg)	\bigg((\mathbf{P}'-\mathbf{k}_\gamma) \mathbf{W}^{(\Lambda)}_{JM}+ C_{lm JM}^{l'm'} B_{J}^{(\Lambda)}\bigg).
\end{split}
\end{equation}
The terms $\frac{\mathbf{P}'\mathbf{k}_\gamma}{M_q}$ and $-\frac{\mathbf{k}_\gamma^2}{2M_q}$ are responsible for the Doppler effect and the quantum recoil, respectively. Further we neglect these contributions, viz., we assume that
\begin{equation}\label{DopplerAndRecoil}
	\frac{\s_c}{M_q}\ll 1,\quad  \frac{k_0}{M_q} \ll 1,\quad k_{0\gamma}\approx\e.
\end{equation}
As the wave packet of the quarkonium center of mass we take
\begin{equation}\label{CMWavePack}
	g(\mathbf{P})=C_{q\bar{q}}e^{-\frac{\mathbf{P}^2}{4\sigma_{c}^2}}e^{-i\mathbf{P}\mathbf{b}_\perp},\quad {|C_{q\bar{q}}|^2=(\sqrt{2\pi}\sigma_{c})^{-3}}.
\end{equation}
Here $\mathbf{b}_\perp:=(b_x,b_y,0)$ is the impact parameter between the propagation axis of the twisted photon and the center of mass of quarkonium. 
We also introduce the notation
\begin{equation}\label{MultipolePhiDependece}
	e^{-iM\phi_k}\Theta_{JM;\lambda}^{(\Lambda)}(\theta_{k}):=\bigg(\mathbf{f}_\lambda(\mathbf{k}) \mathbf{Y}_{JM}^{(\Lambda)*}(\theta_{k},\phi_{k})\bigg) = e^{-iM\phi_k}(-\lambda)^\Lambda \sqrt{\frac{2J+1}{8\pi}} d_{M\lambda}^J(\theta_k),
\end{equation}
where the dependence on $\phi_{k}$ is explicitly shown. The expression for $\Theta_{JM;\lambda}^{(\Lambda)}(\theta_{k})$ is proportional to the small $d$ Wigner functions. Their explicit form is given in expression \eqref{Theta}.
Thus, the amplitude takes the form
\begin{equation}\label{Ampl1}
\begin{split}
	\langle\text{out}|\hat{U}_{\infty,-\infty}|\text{in}\rangle =-\frac{2 iq (2\pi)^4C_{q\bar{q}}}{V\sqrt{2\pi}} \sum_\gamma \frac{\chi_\ga e^{i\lambda\phi_k} }{\sqrt{2 k_{0\ga}}} e^{-\frac{(\spP' - \spk_\ga)^2}{4\sigma_{c}^2}}e^{-i\mathbf{P}'\mathbf{b_\perp}+i\mathbf{k}_\gamma\mathbf{b_\perp}} \delta(\e-k_{0\gamma}) A_{\la}(\spP', \spk_\ga),
\\
	A_{\la}(\spP', \spk_\ga) = \sum_{J\Lambda M}i^{J+\Lambda} e^{-iM\phi_k}\Theta_{JM;\lambda}^{(\Lambda)}(\theta_{k})	\bigg((\mathbf{P}'-\mathbf{k}_\gamma)\mathbf{W}^{(\Lambda)}_{JM}+ C_{lm JM}^{l'm'} B_{J}^{(\Lambda)}\bigg).
\end{split}
\end{equation}

The probability of photoexcitation becomes
 \begin{equation}
 	\begin{split}
 		P_{n'l'm',nlm} =&\frac{2 q^2 (2\pi)^7 {|C_{q\bar{q}}|^2}}{V^2} \sum_{\ga_1,\ga_2} \int \frac{V d\mathbf{P'}}{(2\pi)^3} \frac{\chi_{\ga_1}\chi^*_{\ga_2} e^{i\lambda_1\phi_{k_1}} e^{-i\lambda_2\phi_{k_2}} }{\sqrt{ k_{01}k_{02}}} e^{-\frac{(\spP' - \spk_{1})^2 + (\spP' - \spk_{2})^2}{4\sigma_{c}^2}}e^{i\mathbf{b_\perp}\De_{12}\spk_\perp} \times \\ 
 		\times &\delta(\e-k_{01}) \delta(\e-k_{02}) 
 		 A_{\la_1}(\spP', \spk_{1})A^*_{\la_2}(\spP', \spk_{2}),
 	\end{split}
 \end{equation}
where $\De_{12}\spk = \spk_1-\spk_2 $.
Let us rewrite the argument of the first exponent in the following form
\begin{equation}
	-\frac{(\spP' - \spk_{{1}})^2 + (\spP' - \spk_{{2}})^2}{4\sigma_{c}^2} = -\frac{(\spP'-\frac{\spk_1+\spk_2}{2})^2}{2\s^2_c} - \frac{(\De_{12}\spk)^2}{8\s^2_c}.
\end{equation}
As a result, after changing the variable we obtain
\begin{equation}
	\begin{split}
		P_{n'l'm',nlm} =&\frac{2 q^2 (2\pi)^4 {|C_{q\bar{q}}|^2}}{V\e} \sum_{\ga_1,\ga_2} \int d\mathbf{P'} \chi_{\ga_1}\chi_{\ga_2}^{{*}}  e^{i\lambda_1\phi_{k_1}} e^{-i\lambda_2\phi_{k_2}} e^{-\frac{(\spP')^2}{2\sigma_{c}^2}} e^{- \frac{\De_{12}\spk^2}{8\s^2_c}} e^{i\mathbf{b_\perp}\De_{12}\spk_\perp} \times \\ 
		\times &\delta(\e-k_{01}) \delta(\e-k_{02}) 
		A_{\la_1}(\spP'  + \frac{\spk_1+\spk_2}{2}, \spk_{1})A^*_{\la_2}(\spP' + \frac{\spk_1+\spk_2}{2}, \spk_{2} ).
	\end{split}
\end{equation}

The integral over $ \spP' $ is Gaussian and can be evaluated
\begin{equation}
\begin{split}
P^{(Tw)}_{n'l'm',nlm} =&\frac{2 q^2 (2\pi)^4}{V\e} \sum_{J_1 J_2 } \sum_{M_1 M_2} \sum_{\La_1\La_2} i^{J_1+\La_1} (-i)^{J_2+\La_2}\sum_{\ga_1,\ga_2} \chi_{\ga_1}\chi_{\ga_2}^{{*}}  e^{i\lambda_1\phi_{k_1}} e^{-i\lambda_2\phi_{k_2}}  e^{- \frac{\De_{12}\spk^2}{8\s^2_c}} e^{i\mathbf{b_\perp}\De_{12}\spk_\perp} \times \\ 
\times& \delta(\e-k_{01}) \delta(\e-k_{02})e^{-i M_1 \phi_{k_1}}e^{i M_2 \phi_{k_2}} \Theta_{J_1M_1;\la_1}^{(\La_1)}(\theta_{k_1}) \Theta_{J_2M_2;\la_2}^{*(\La_2)}(\theta_{k_2}) Q^{(\La_1)(\La_2)}_{J_1M_1J_2M_2}(\De_{12}\spk),\\
\end{split}
\end{equation}
where, for the sake of brevity, we have introduced the notation
\begin{equation}
\begin{split}
Q^{(\La_1)(\La_2)}_{J_1M_1J_2M_2}&(\De_{12}\spk) = \big[C_{lm J_1M_1}^{l'm'} C_{lm J_2M_2}^{l'm'}B_{J_1}^{(\La_1)}B_{J_2}^{*(\La_2)} +{\s_c^2} \mathbf{W}^{(\La_1)}_{J_1M_1}\mathbf{W}^{*(\La_2)}_{J_2M_2} - {\frac{1}{4} (\De_{12}\spk \cdot \mathbf{W}^{(\La_1)}_{J_1M_1}) (\De_{12}\spk\cdot\mathbf{W}^{*(\La_2)}_{J_2M_2})} + \\ 
+& \frac{\De_{12}\spk}{2}\big(C_{lm J_1M_1}^{l'm'}B_{J_1}^{(\La_1)}\mathbf{W}^{*(\La_2)}_{J_2} -  C_{lm J_2M_2}^{l'm'}B_{J_2}^{*(\La_2)} \mathbf{W}^{(\La_1)}_{J_1M_2} \big)  \big].
\end{split}
\end{equation}

Consider the wave packet of a twisted photon in the form
\begin{equation}\label{BesselWavePack}
	\chi_\gamma = C_\gamma k_\perp^{{|l_\gamma|}} e^{-\frac{(k_\perp^2-(k_\perp^0)^2)^2}{4\sigma_{\perp}^4}}e^{-\frac{(k_3-k_3^0)^2}{4\sigma_3^2}}e^{il_\gamma\phi_k}\delta_{\lambda\lambda_0}, \quad \gamma = (\lambda,\mathbf{k}),\quad \sum_{\gamma}\equiv \sum_{\lambda=\pm 1}\int_0^{\infty} \frac{V k_\perp d k_\perp}{(2\pi)^3} \int_{-\infty}^{\infty} d k_3 \int_{0}^{2\pi} d\phi_{k},
\end{equation}
where $C_\ga$ is the normalization factor, $l_\gamma:=m_\gamma - \lambda$. Further we omit for the sake of brevity the limits of integration with respect to $ k_\perp, k_3, \phi_k$. With $\s_\perp$ and $\s_3$ tending to zero, this state becomes a Bessel state of a twisted photon. Notice that the dependence of the wave packet profile on $k_\perp$ must be of the form $k_\perp^{{|l_\gamma|}}g_{m_\ga}(k_\perp^2)$, where $g_{m_\ga}(k_\perp^2)$ is some smooth function provided that the function of $\spk$ defining the shape of the wave packet is smooth at $k_\perp=0$. The normalization constant is found from condition \eqref{norm_conds}, which is written as
\begin{equation} 
	|C_\ga|^2\int\frac{Vd\spk}{(2\pi)^3} k_{\perp}^{2{|l_\ga|}} e^{-\frac{[k_{\perp}^2-(k_\perp^0)^2]^2}{2 \sigma_\perp^4}}
	e^{-\frac{(k_{3}-k_3^0)^2}{2 \sigma_3^2}}=1.
\end{equation}
Then the probability of photoexcitation of quarkonium by a twisted photon is reduced to
\begin{equation}\label{PrTwIntermediate}
	\begin{split}
		P^{(Tw)}_{n'l'm',nlm} =&\frac{2 V q^2 |C_\ga|^2}{(2\pi)^2\e} \int d k_{\perp 1} d k_{\perp 2} d k_{31}  d k_{32} d\phi_{k_1} d\phi_{k_2}  k_{\perp 1}^{{|l_\gamma| + 1}}k_{\perp 2}^{{|l_\gamma| + 1}} e^{- \frac{\De_{12}\spk^2}{8\s^2_c}} e^{i\mathbf{b_\perp}\De_{12}\spk_\perp} \times \\ 
		\times&   e^{-\frac{(k_{\perp1}^2-(k_\perp^0)^2)^2}{4\sigma_{\perp}^4}}e^{-\frac{(k_{\perp2}^2-(k_\perp^0)^2)^2}{4\sigma_{\perp}^4}} e^{-\frac{(k_{31}-k_3^0)^2}{4\sigma_3^2}} e^{-\frac{(k_{32}-k_3^0)^2}{4\sigma_3^2}} e^{im_\gamma (\phi_{k_1} - \phi_{k_2})} \delta(\e-k_{01}) \times\\
		\times & \delta(\e-k_{02}) \sum_{J_1 J_2 } \sum_{M_1 M_2} \sum_{\La_1\La_2} i^{J_1+\La_1} (-i)^{J_2+\La_2} 
		 e^{-i M_1 \phi_{k_1}}e^{i M_2 \phi_{k_2}} \Theta_{J_1M_1;\la_0}^{(\La_1)}(\theta_{k_1})  \times\\ 
		 \times & \Theta_{J_2M_2;\la_0}^{*(\La_2)}(\theta_{k_2}) Q^{(\La_1)(\La_2)}_{J_1M_1J_2M_2}(\De_{12}\spk),
	\end{split}
\end{equation}
The integrals with respect to $k_3$ can be evaluated using the delta function, given that
\begin{equation}
	\delta(\e-k_{0\gamma}) = \frac{k_{0\gamma}}{\tilde{k}_3(k_\perp)} [\delta(k_3-\tilde{k}_3(k_\perp)) + \delta(k_3+\tilde{k}_3(k_\perp))], \quad \tilde{k}_3(k_\perp) :=\sqrt{\e^2-k_\perp^2}.
\end{equation}
Notice, that the contribution stemming from the second $\de$-function in this expression is much smaller than the contribution from the first one due to the Gaussians with respect to $k_3$ in \eqref{PrTwIntermediate}. Thus we take
\begin{equation}
	\delta(\e-k_{0\gamma}) \approx \frac{k_{0\gamma}}{\tilde{k}_3(k_\perp)} \delta(k_3-\tilde{k}_3(k_\perp)).
\end{equation}

The integral over $k_\perp$ can be calculated perturbatively with respect to $\sigma_{\perp}$. To this aim, the following conditions must be fulfilled (see also \cite{KS2024})
\begin{equation}\label{main_cond}
	\frac{k^0_\perp\s_\perp}{(k^0_3)^2} \ll 1,\quad \frac{k^0_\perp\s_\perp}{\s_c^2} \ll 1, \quad \frac{k^0_\perp \s_\perp \s_3}{\s_c^2 k_3^0} \ll 1,\quad\s_\perp b_\perp \ll 1, \quad \frac{(|m_\ga| + J + 1)\s_\perp}{k^0_\perp} \ll 1.
\end{equation}
One can estimate $J = |m_\ga| $, so the last condition can be replaced by 
\begin{equation}
	\dfrac{2(|m_\ga| + 1)\s_\perp}{k^0_\perp} \ll 1.
\end{equation}
Therefore,
\begin{equation}
	\begin{split}
		P^{(Tw)}_{n'l'm',nlm} =&\frac{ 2 V \e q^2 |C_\ga|^2 \s_\perp^4( k^0_{\perp})^{{2|l_\gamma|}}}{4 \pi (\e^2-(k_\perp^0)^2)}  e^{-\frac{(\tilde{k}_{3}(k_\perp^0)-k_3^0)^2}{2\sigma_3^2}} \int  d\phi_{k_1} d\phi_{k_2} e^{- \frac{\De_{12}\spk_\perp^2}{8\s^2_c}} e^{i\mathbf{b_\perp}\De_{12}\spk_\perp} e^{im_\gamma (\phi_{k_1} - \phi_{k_2})}  \times \\ 
		\times& 
		\sum_{J_1 J_2 } \sum_{M_1 M_2} \sum_{\La_1\La_2} i^{J_1 - J_2 + \La_1 - \La_2} e^{-i M_1 \phi_{k_1} + i M_2 \phi_{k_2}} \Theta_{J_1M_1;\la_0}^{(\La_1)}(\theta_k^{0}) \Theta_{J_2M_2;\la_0}^{*(\La_2)}(\theta_k^{0}) \times \\ \times & Q^{(\La_1)(\La_2)}_{J_1M_1J_2M_2}(\De_{12}\spk_\perp),
	\end{split}
\end{equation}
where we set $ k_{31} = k_{32} = \tilde{k}_{3}(k_\perp^0) = \sqrt{\e^2-(k_\perp^0)^2} $, $ k_{\perp1} = k_{\perp2} = k_\perp^0$, and $\cos\theta_k^0=\tilde{k}_{3}(k_\perp^0)/\e$. In addition, in the leading order with respect to $\s_\perp$ 
\begin{equation}
	|C_\ga|^2 = \frac{{(2\pi)^3}}{2\pi^2 V \s_3 \s_\perp^2(k_\perp^0)^{{2|m_\gamma-\lambda_0|}}}.
\end{equation}
Thus the probability of photoexcitation takes the form
\begin{equation}\label{PrTw1}
\begin{split}
	P^{(Tw)}_{n'l'm',nlm} &={\frac{2 \e q^2 \s_\perp^2}{\s_3 \tilde{k}^2_{3}(k_\perp^0)}}e^{-\frac{( \tilde{k}_{3}(k_\perp^0)-k_3^0)^2}{2\sigma_3^2}}\sum_{J_1 J_2 } \sum_{M_1 M_2} \sum_{\La_1\La_2} i^{J_1+\La_1} (-i)^{J_2+\La_2}  \int  d\phi_{k_1} d\phi_{k_2} e^{- \frac{\De_{12}\spk_\perp^2}{8\s^2_c}} e^{i\mathbf{b_\perp}\De_{12}\spk_\perp} e^{im_\gamma (\phi_{k_1} - \phi_{k_2})} \times \\ 
	&\times e^{-i M_1 \phi_{k_1}}e^{i M_2 \phi_{k_2}} \Theta_{J_1M_1;\la_0}^{(\La_1)}(\theta_{k}^0) \Theta_{J_2M_2;\la_0}^{*(\La_2)}(\theta_{k}^0)\Big[C_{lm J_1M_1}^{l'm'} C_{lm J_2M_2}^{l'm'}B_{J_1}^{(\La_1)}B_{J_2}^{*(\La_2)} + {\s_c^2} \mathbf{W}^{(\La_1)}_{J_1M_1}\mathbf{W}^{*(\La_2)}_{J_2M_2} -\\
	&-  {\frac{1}{4} (\De_{12}\spk_\perp \cdot \mathbf{W}^{(\La_1)}_{J_1M_1}) (\De_{12}\spk_\perp\cdot\mathbf{W}^{*(\La_2)}_{J_2M_2})} + \frac{\De_{12}\spk_\perp}{2}\big(C_{lm J_1M_1}^{l'm'} B_{J_1}^{(\La_1)}\mathbf{W}^{*(\La_2)}_{J_2M_2} - C_{lm J_2M_2}^{l'm'}B_{J_2}^{*(\La_2)} \mathbf{W}^{(\La_1)}_{J_1M_1} \big)  \Big].
\end{split}
\end{equation}
From the explicit form of $ B^{(\La)}_{J} $ and $ \mathbf{W}^{(\La)}_{JM} $ and the conditions \eqref{main_cond} , we can see that the first two terms in the square brackets of \eqref{PrTw1} are greater than the terms on the last line. More precisely, the terms in the square brackets are of order $ \e^2 $, $ \s_c^2 $, $ \e^2 (n_\perp^0)^2 $, and $ \e^2 n_\perp^0 $, respectively, where $ n_\perp^0 = k_\perp^0/\e $. Thus, the contributions of the largest terms in the paraxial limit $ n_\perp^0 \ll 1$ are independent of $ \phi_k $. In fact, after summing over magnetic quantum numbers, which will be performed below, the terms in the last round brackets in \eqref{PrTw1} turn out to be zero. Thus we only need to calculate the integrals with respect to angular variables $\phi_{k_{1,2}}$. This calculations are presented in the App. \ref{phiint}. As a result, on the paraxial approximation we come to 
\begin{equation}\label{PrTw2}
	\begin{split}
		P^{(Tw)}_{n'l'm',nlm} &={\frac{8 \pi^2 \e q^2 \s_\perp^2}{\s_3 \tilde{k}^2_{3}(k_\perp^0) }}e^{-\frac{( k_\perp^0)^2}{4\s_c^2}} e^{-\frac{( \tilde{k}_{3}(k_\perp^0)-k_3^0)^2}{2\sigma_3^2}} \sum_{J_1 J_2 } \sum_{M_1 M_2} \sum_{\La_1\La_2} \sum_{n=-\infty}^{\infty} i^{J_1+\La_1 + M_1 - J_2 - \La_2 - M_2} e^{-i(M_1-M_2)\phi_b} \times\\
		& \times \Theta_{J_1M_1;\la_0}^{(\La_1)}(\theta_{k}^0) \Theta_{J_2M_2;\la_0}^{*(\La_2)}(\theta_{k}^0) 
		I_n\bigg(\frac{( k_\perp^0)^2}{4\s_c^2}\bigg)J_{M_1-m_\ga-n}(k_\perp^0b_\perp)J_{M_2-m_\ga-n}(k_\perp^0b_\perp)  \times\\
		& \times \big[C_{lm J_1M_1}^{l'm'} C_{lm J_2M_2}^{l'm'}B_{J_1}^{(\La_1)}B_{J_2}^{*(\La_2)} + {\s_c^2} \mathbf{W}^{(\La_1)}_{J_1M_1}\mathbf{W}^{*(\La_2)}_{J_2M_2} \big].
	\end{split}
\end{equation}

Usually the initial states of quarkonia are mixed with respect to the quantum number $ m $, the states corresponding to different $m$ appear with equal probabilities, and the projection $ m' $ is not measured. In this regard it is reasonable to consider the probability averaged over $ m $ and summed over $ m' $
\begin{equation}\label{PrmSum}
	P^{(Tw)}_{n'l',nl} := \frac{1}{2l+1}\sum_{m=-l}^{l}\sum_{m'=-l'}^{l'} P^{(Tw)}_{n'l'm',nlm}.
\end{equation}
We see from the explicit expression for $ \mathbf{W}^{(\La)}_{JM} $ that its dependence of $ m $ and $ m' $ is specified by the Clebsch-Gordan coefficients. In order to evaluate the sum in \eqref{PrmSum}, we use the relation
\begin{equation}
	\sum_{m=-l}^{l}\sum_{m'=-l'}^{l'}  C_{lm J_1M_1}^{l'm'} C_{lm J_2M_2}^{l'm'}  = \frac{2l'+1}{2J_1+1} \de_{J_1J_2} \de_{M_1M_2}. 
\end{equation}
Then the terms in \eqref{PrTw2} with $\La_1 \neq \La_2 $ vanish. As a result, we obtain
\begin{equation}\label{PrTwSumm}
\begin{split}
P^{(Tw)}_{n'l',nl} =&\frac{8 \pi^2 \e q^2 \s_\perp^2}{\s_3 \tilde{k}^2_{3}(k_\perp^0)}\frac{2l'+1}{2l+1} e^{-\frac{( k_\perp^0)^2}{4\s_c^2}} e^{-\frac{( \tilde{k}_{3}(k_\perp^0)-k_3^0)^2}{2\sigma_3^2}} \sum_{J=1}^{\infty} \sum_{M = -J}^{J} \sum_{\La = 0,1} \sum_{n=-\infty}^{\infty}\frac{1}{2J+1} I_n\bigg(\frac{( k_\perp^0)^2}{4\s_c^2}\bigg)J^2_{M-m_\ga-n}(k_\perp^0b_\perp) \times \\
&\times  |\Theta_{JM;\la_0}^{(\La)}(\theta_{k}^0)|^2  \big[|B_{J}^{(\La)}|^2 + {\s_c^2} |W^{(\La)}_{JM}|^2 \big],
\end{split}
\end{equation}
where $ |W^{(\La)}_{JM}|^2 $ is defined in \eqref{W_JMFinal}.

It is clear, that for $k_\perp^0 \ll 2  \s_c $, the modified Bessel function
\begin{equation}
	I_n\big(\frac{( k_\perp^0)^2}{4\s_c^2}\big)\approx \delta_{n,0}.
\end{equation} 
If in addition the condition $ k_\perp^0b_\perp \ll 1$ is satisfied, then the photoexcitation probability takes the form
\begin{equation}\label{PrTwSummSiml}
	\begin{split}
		P^{(Tw)}_{n'l',nl} =&\frac{8 \pi^2 \e q^2 \s_\perp^2}{\s_3 \tilde{k}^2_{3}(k_\perp^0)}\frac{2l'+1}{2l+1} e^{-\frac{( k_\perp^0)^2}{4\s_c^2}} e^{-\frac{( \tilde{k}_{3}(k_\perp^0)-k_3^0)^2}{2\sigma_3^2}}\sum_{J=\max(|m_\ga|,1)}^{\infty} \sum_{\La = 0,1} \frac{1}{2J+1} |\Theta_{Jm_\ga;\la_0}^{(\La)}(\theta_{k}^0)|^2  \big[|B_{J}^{(\La)}|^2 + {\s_c^2} |W^{(\La)}_{Jm_\ga}|^2 \big].
	\end{split}
\end{equation}
Thus, for $|m_\ga| > 1$ the multipole transitions with $ J \geq |m_\ga|$ are realized. The transitions with $J < |m_\ga|$ have a higher order of smallness with respect to the the parameter $ k_\perp^0b_\perp$. Moreover, analyzing the explicit form of the matrix elements in the square brackets, one can see that for low-energy transitions, i.e., $ \e r_B \ll 1 $, $r_B$ being the corresponding Bohr radius, and when $|l'-l| > 1$, the leading contribution to the sum comes from the term containing $B^{(1)}_{|l'-l|}$. Therefore we conclude that for a fixed $l$ the leading contribution corresponds to the electric transition $E|m_\gamma|$ inducing the transition $l \rightarrow l' = l + m_\ga$. For $|m_\ga| \leq 1$ the leading contribution comes from the standard dipole transition. 
Notice, that the  probability of photoexcitation of quarkonia by twisted photons \eqref{PrTwSummSiml} is quite similar to the probability of photoabsorption of a twisted photon in \cite{KS2024}. The key differences between these probabilities stem from the usage of Wigner $d$-matrices in \cite{KS2024} and from the accounting for the center of mass dynamics in the present work.

\section{Numerical simulation}\label{SectionNumerical}

Typical transition energies for quarkonium are of the order of 100 MeV \cite{Kwong1988} that significantly complicates the studies of photoexcitation of quarkonia. However, in the considered model \eqref{CornellPotential} taking into account the relativistic corrections \eqref{BreitPotential}, the charmonium system can undergo a transition $1^{3}G_3 \rightarrow 3^{3}P_0$ with an exceptionally low energy $\e= E_{1^{3}G_3 \rightarrow 3^{3}P_0} \approx 0.756$ MeV. Such a transition is octupole, i.e.,  $|l'-l|=3$. Therefore, this transition will be excited by a twisted photon with $m_\gamma=3$. We consider this peculiar transition as an example.

We numerically simulate formula \eqref{PrTw1} summed with respect to $m'$ and averaged with respect to $m$. 
The parameters of the charmonium system are chosen in accordance with \cite{eiglsperger2007}: $m_q=1.2185$ GeV, $\alpha_s =0.29$, $\s=1.306$ GeV/fm. The parameters of the wave packets are chosen as follows. The average values of longitudinal and transverse momenta of photons $k_3^0=\e\sqrt{1-(n^0_\perp)^2} \approx 0.753$ MeV, $k_\perp^0 = n^0_\perp \e \approx 0.0756$ MeV, where $n^0_\perp=0.1$. We estimate the photon beam dispersion as $\approx 1\%$ of the average momentum values, viz., $\s_3 = 10$ keV and $\s_\perp = 1$ keV. These values of the wave packet parameters can be achieved, for example, in inverse Compton scattering \cite{Jentshura2011_1, Jentshura2011_2, BKL2019, Ivanov2022, Guo2023}. The value of momentum dispersion of the quarkonium center of mass is limited from above by condition \eqref{DopplerAndRecoil}. We estimate it as $\s_c=0.1$ GeV (see also \cite{adler2003, LHCb2023, Fleming2020}). We choose the impact parameter to be $b_\perp = 100 \; \text{fm} $. Such a value is necessary for expression \eqref{PrTwSummSiml} to be true. It can, however, be increased to angstroms, or even microns by abandoning the simplified expression for the probability \eqref{PrTwSummSiml} and employing a more general one \eqref{PrTwSumm}. The parameters are selected to maximize the probability of photoexcitation while remaining consistent with the estimates \eqref{main_cond} and \eqref{DopplerAndRecoil}.

\begin{figure}[tp]
	\centering
	\includegraphics*[width=0.5\linewidth]{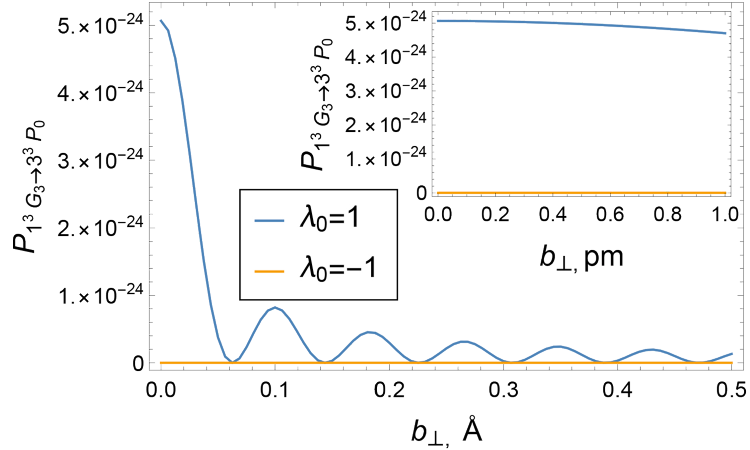}
	\caption{{\footnotesize
			Photoexcitation probability for the charmonium transition $1^{3}G_3 \rightarrow 3^{3}P_0$ by a twisted photon with $m_\gamma = 3$ and the energy $ 0.756 $ MeV for the different circular polarizations with respect to the impact parameter $b_\perp$. The parameters are: $m_q=1.2185$ GeV, $\alpha_s =0.29$, $\s=1.306$ GeV/fm, $\s_c=0.1$ GeV,  $k_3^0=0.753$ MeV, $k_\perp^0 = 0.0756$ MeV, $\s_3 = 10$ keV, and $\s_\perp = 1$ keV. }}
	\label{Figbp}
\end{figure}

\begin{figure}[tp]
	\centering
	\includegraphics*[width=0.49\linewidth]{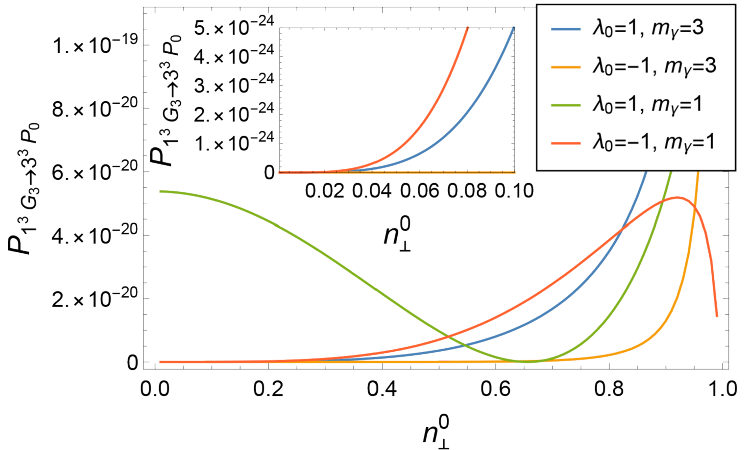}
	\includegraphics*[width=0.5\linewidth]{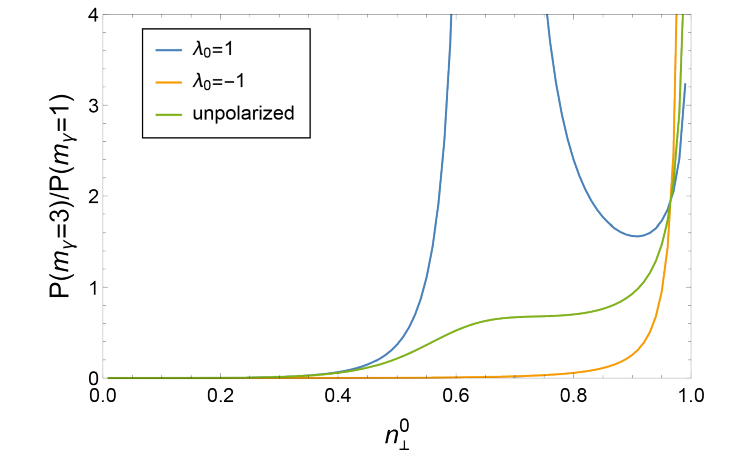}
	\caption{{\footnotesize
			Comparison of photoexcitation probabilities for the charmonium transition $1^{3}G_3 \rightarrow 3^{3}P_0$ by photons of the same energy $ 0.756 $ MeV with $m_\gamma = 3$ and $m_\gamma = 1$ and different polarizations with respect to $n^{0}_\perp$. On the left panel: The absolute values of the probabilities. On the right panel: The ratio of the probability of photoexcitation by a twisted photon ($m_\gamma = 3$) to the probability of photoexcitation by a plane-wave photon ($m_\gamma = 1$). The parameters are: $m_q=1.2185$ GeV, $\alpha_s =0.29$, $\s=1.306$ GeV/fm, $\s_c=0.1$ GeV, $\s_3 = 10$ keV, $\s_\perp = 1$ keV, and $b_\perp = 100 \; \text{fm} $.}}
	\label{ProbRatio}
\end{figure}

The dependence of the inclusive probability on $b_\perp$ for the different circular polarizations of incident photons is shown in Fig. \ref{Figbp}. It is also clear from this Figure, that the probability is  sensitive to the helicity of the incident photon. Furthermore, it should be noticed, that the impact parameter can be increased up to several tenth of pm without significant decrease in the transition probability. As can be seen from both Figs.\ref{Figbp} and \ref{ProbRatio}, the transition probability for the chosen parameters (and the photon helicity $\la_0=1$) turns out to be of the order of  $P_{1^{3}G_3 \rightarrow 3^{3}P_0}\sim 10^{-24}$. This is expected for an octupole transition in a binary system with $r_B\approx 1 $ fm. It is known that the multipole transitions can be excited by a plane-wave photon. In the paraxial limit $n_\perp \ll 1$ the probability of this octupole transition induced by a plane-wave appears to be several orders of magnitude greater then the probability of the same transition induced by an appropriate twisted photon. Figure \ref{ProbRatio} reveals regimes where twisted photons can outperform plane-wave ones in exciting transitions suppressed in the dipole approximation. Therefore it is possible to enhance non-dipole excitation rate, in particularly by tuning $n_\perp$, the beam geometry, or the polarization.

\section{Conclusion}\label{SectionConclusion}
Let us recapitulate the results of this study. In the framework of potential non-relativistic QCD, we have investigated the process of excitation of multipole transitions in the quarkonium system by a photon with a nonzero projection of the orbital angular momentum described by the regularized Bessel state \eqref{BesselWavePack}. We have derived the explicit expressions for the amplitude and probability of photoexcitation of quarkonium by a twisted photon taking into account the dynamics of the quarkonium center of mass. For this purpose, we have introduced a center-of-mass wave packet $Q\bar{Q}$ \eqref{CMWavePack} characterized by the dispersion in momentum space $\sigma_{c}$ and the impact parameter $b_\perp$. We have obtained the selection rules for transitions induced by twisted photons for the contributions to the probability amplitude coming from the relative motion of quarks and from the dynamics of the center of mass. It is shown that the leading contribution to the photoexcitation probability comes from the multipole $j=|m_\gamma|$ and $m_\gamma=l'-l$ in the long-wave approximation, which expresses the angular momentum conservation law. However, the fulfillment of these selection rules imposes constraints on the parameters of the wave packet of the quarkonium center of mass, since, in general, the angular momentum can be transferred to the center of mass (see, e.g., for molecules \cite{Maslov2024} and for excitons \cite{GrBhSeHa22}). The transition $j=|m_\gamma|$ can also be excited by a plane-wave photon, but in the case of the absorption of a twisted photon with $|m_\gamma|>1$, the multipole transitions with $j=|m_\gamma|$ do not overlap with the transitions of lower multipolarity and can be studied separately.

As an example, we have considered the octupole transition $1^{3}G_3 \rightarrow 3^{3}P_0$ in the charmonium system. We do not specify the type of this transition because, as follows from the parity selection rules \eqref{BSelRul}, \eqref{WSelRul}, the leading contribution to the photoabsorption amplitude from the relative motion of the quarks is characterized as electric $E3$ and from the motion of the center of mass as magnetic $M3$. Such a transition has a very low energy of $ 0.756$ MeV as compared to the typical values of transition energies in quarkonia system of tens or even hundreds of MeV. The twisted photons with such energies can be created by using the inverse Compton scattering \cite{Jentshura2011_1, Jentshura2011_2, BKL2019, Ivanov2022, Guo2023} or channeling radiation \cite{ABKT2018,BKT2022}. The parameters of the wave packets have been chosen to maximize the probability, to satisfy the estimates \eqref{main_cond}, \eqref{DopplerAndRecoil}, and to match roughly the experimentally achievable values. As expected, we obtained a rather small value for the photoexcitation probability on the order of $10^{-24}$ in the paraxial regime. We have also compared the probabilities of excitation of this multipole transition by plane-wave and twisted photons with different polarizations (see Fig. \ref{ProbRatio}). The probability can be increased by abandoning the paraxial approximation, see Fig. \ref{ProbRatio}. In the example we have considered, there is also the problem of creating $1^{3}G_3$ state that has not been reliably observed in experiments, or at least the observed candidates do not fit the potential model. Another possible hindrance for the realistic experiment with the simulated transition is the relatively short expected lifetime of the $1^{3}G_3$ charmonium state.

The developed theory allows us to consider less exotic transitions. However, the potential experiments will require a source of hard twisted photons with much higher energy. When such sources appear, the spectra of multipole transitions in quarkonium will allow us to impose new constraints on the parameters of the existing effective potentials and to find a more accurate model describing quarkonium by solving the inverse scattering problem. The developed formalism can be applied to other binary quantum systems such as positronium, excitons, or Cooper pairs.

\paragraph{Acknowledgments.}

We thank Dr P.O. Kazinski for the problem statement and for many insightful discussions and constructive feedback concerning this work.

\appendix

\section{Matrix element $\kappa_\alpha^-$}\label{SmallKappaEvaluation}
We need to calculate the matrix element
\begin{equation}
	\kappa_\ga  ^{-}= - m_q^{-1}\langle\text{out}| \hat{c}_\ga   \big(\Psi_{\ga   i}(\mathbf{R}-\frac{1}{2}\mathbf{r}) +\Psi_{\ga   i}(\mathbf{R}+\frac{1}{2}\mathbf{r}) \big)p_i |\text{in}\rangle.
\end{equation}
It is clear that it is sufficient to evaluate the matrix element
\begin{equation}
	\begin{split}
		\langle\text{out}| \hat{c}_\ga \Psi_{\ga i}(\mathbf{R}-b\mathbf{r}) \hat{p}_i |\text{in}\rangle=\frac{\chi_\ga e^{i\lambda\phi_k}}{\sqrt{2 k_{0\ga}}}\int\frac{ d\mathbf{P}}{(2\pi)^{3/2}} g(\mathbf{P})\langle \mathbf{P}'|e^{i\mathbf{R}\mathbf{k}}|\mathbf{P}\rangle \langle n'l'm'|f_{\lambda i}(\mathbf{k})e^{-ib\mathbf{r}\mathbf{k}}\hat{p}_i |nlm\rangle.
	\end{split}
\end{equation}
Let us write the matrix element associated with the relative motion in a symmetric form, taking into account the Coulomb gauge
\begin{equation}
	\langle n'l'm'|f_{\lambda i}(\mathbf{k})e^{-ib\mathbf{r}\mathbf{k}}\hat{p}_i |nlm\rangle = \frac{1}{2} \langle n'l'm'|\hat{p}_i f_{\la i}(\mathbf{k})e^{-ib\mathbf{r}\mathbf{k}} + f_{\lambda i}(\mathbf{k})e^{-ib\mathbf{r}\mathbf{k}}\hat{p}_i |nlm\rangle.
\end{equation}
In this expression the momentum operator in the first term acts to the left and in the second term acts to the right. Then we obtain
\begin{multline}\label{matr_elem_i_}
	-\frac{i}{2} \int_0^\infty drr^2 \int d\theta  d\vf \sin\theta f_{\lambda i}(\mathbf{k})e^{-ib\mathbf{r}\mathbf{k}} \big[R^*_{n'L'J'}(r)Y^*_{L'M'_L}(\theta,\vf) \nabla(R_{nLJ}(r)Y_{LM_L}(\theta,\vf)) - \\
	- R_{nLJ}(r)Y_{LM_L}(\theta,\vf)
	\nabla(R^*_{n'L'J'}(r)Y^*_{L'M'_L}(\theta,\vf))\big].
\end{multline}
A plane wave is expanded in terms of spherical vectors as \cite{varshalovich1988}
\begin{equation}\label{multipole_expension}
	\mathbf{f}_\lambda(\mathbf{k})e^{-ib\mathbf{k}\mathbf{r}}= 4\pi \sum_{JM\La}(-i)^\La i^J\bigg(\mathbf{f}_\la(\mathbf{k}) \mathbf{Y}_{JM}^{(\La)*}(\theta_{k},\phi_{k})\bigg) \bs\psi^\La_{JM}(k_0,-b\spr),
\end{equation}
where the mode functions have the form
\begin{equation}
	\bs\psi^\La_{JM}(k_0,\spr)=
	\left\{
	\begin{array}{ll}
		j_J(k_0r)\mathbf{Y}^{(0)}_{JM}(\theta,\vf), & \hbox{$\La=0$;} \\
		(j'_{J}(k_0r) + \frac{j_{J}(k_0r)}{k_0 r})\mathbf{Y}^{(1)}_{JM}(\theta,\vf) +\sqrt{J(J+1)}\frac{j_{J}(k_0r)}{k_0r}
		\mathbf{Y}^{(-1)}_{JM}(\theta,\vf), & \hbox{$\La=1$},
	\end{array}
	\right.
\end{equation}
where
\begin{equation}
	\mathbf{Y}^{(1)}_{J M}(\theta,\vf) = -i [\mathbf{n}, \mathbf{Y}^{(0)}_{J M}(\theta,\vf)],\qquad \mathbf{Y}^{(0)}_{J M}(\theta,\vf) = \mathbf{Y}^{J}_{JM}(\theta,\vf),\qquad \mathbf{Y}^{(-1)}_{JM}(\theta,\vf) = \mathbf{n} Y_{JM}(\theta,\vf),
\end{equation}
and $\mathbf{n} $ is the unit vector along the vector $\mathbf{r}$. Taking into account such a decomposition, the amplitude can be cast into the form
\begin{equation}
\begin{split}
	\kappa_\ga  ^{-}&= \frac{4 \pi i}{2} \frac{\chi_\ga e^{i\lambda\phi_k}}{\sqrt{2 k_{0\ga}}} \int\frac{ d\mathbf{P}}{(2\pi)^{3/2}} g(\mathbf{P})\langle \mathbf{P}'|e^{i\mathbf{R}\mathbf{k}}|\mathbf{P}\rangle \sum_{JM\La} (1 + (-1)^{J+\La})(-i)^\La i^J\bigg(\mathbf{f}_\la(\mathbf{k}) \mathbf{Y}_{JM}^{(\La)*}(\theta_{k},\phi_{k})\bigg) \times \\ 
	&\times \int_0^\infty drr^2 \int d\theta  d\vf  \sin\theta \bs \psi^\La_{JM}(k_0,\spr/2) \big[R^*_{n'L'J'}(r)Y^*_{L'M'_L}(\theta,\vf) \nabla(R_{nLJ}(r)Y_{LM_L}(\theta,\vf)) -\\
	&- R_{nLJ}(r)Y_{LM_L}(\theta,\vf)	\nabla(R^*_{n'L'J'}(r)Y^*_{L'M'_L}(\theta,\vf))\big].
\end{split}
\end{equation}
In calculating the integrals, we use the properties of the Clebsch-Gordon coefficients, the spherical vectors, and the gradient formula \cite{varshalovich1988}
\begin{equation}
	\begin{split}
		\nabla\big(R_{nl}(r)Y_{lm}\big)= R'_{nl}(r)\mathbf{Y}_{lm}^{(-1)}(\theta,\phi) + \sqrt{l(l+1)}\frac{R_{nl}(r)}{r}\mathbf{Y}_{lm}^{(1)}(\theta,\phi).
	\end{split}
\end{equation}
Then
\begin{equation}\label{MultipoleExpansionForSmallKappa}
	\begin{split}
		\kappa_\gamma^{-} = \frac{4\pi \chi_\ga e^{i\lambda\phi_k}}{  \sqrt{2k_{0\gamma}}}\int\frac{ d\mathbf{P}}{(2\pi)^{3/2}} g(\mathbf{P}) \langle \mathbf{P}'|e^{i\mathbf{R}\mathbf{k}}|\mathbf{P}\rangle  \sum_{J\Lambda M} i^{J+\La}\bigg(\mathbf{f}_\lambda(\mathbf{k}) \mathbf{Y}_{JM}^{(\Lambda)*}(\theta_{k},\phi_{k})\bigg) C_{lm JM}^{l'm'}B_{J}^{(\Lambda)},
	\end{split}
\end{equation}
where
\begin{equation}\label{B_JM^La}
	\begin{split}
		&B_{J}^{(0)}  = -i \frac{1 + (-1)^{J}}{m_q}  C_{l0 J0}^{l'-1 ,0} \sixj{J}{J}{1}{l'-1}{l'}{l} (2J+1)\sqrt{\frac{l'(2l'+1)(2l+1)}{4 \pi}} R_{n'l'nl}^J, \\
		&B_{J}^{(1)}  =  -i\frac{1 - (-1)^{J}}{2m_q} C_{l0 J0}^{l' 0}\sqrt{\frac{(J+1)(2J+1)(2l+1)}{4 \pi J(2l'+1)}} \bigg\{ \frac{2J}{k_0} \int_0^\infty dr r j_J\bigg(\frac{k_0r}{2}\bigg) W(r)+ \\
		&\quad \quad \quad \quad \quad \quad \quad \quad \quad\quad\quad\quad\quad + \frac{l(l+1) - l'(l'+1)}{(J+1)(2J+1)}  \big[ (J+1) R^{J-1}_{n'l'nl} - J  R^{J+1}_{n'l'nl}\big]\bigg\},\\
		&R_{n'l'nl}^J := \int_0^\infty dr r j_J\bigg(\frac{k_0r}{2}\bigg)R_{nl}(r) R^*_{n'l'}(r),
	\end{split}
\end{equation}
and $W(r) =  R'_{nl}(r) R^*_{n'l'}(r) - R_{nl}(r) R'^*_{n'l'}(r) $ is the Wronskian of solutions of the radial Schr{\"o}dinger equation for the Hamiltonian \eqref{RadShr}. The selection rules \eqref{BSelRul} follow from expressions \eqref{MultipoleExpansionForSmallKappa} and \eqref{B_JM^La}.

The sum in the expression \eqref{MultipoleExpansionForSmallKappa} is a multipole expansion for the transition amplitude. The terms with $\Lambda=1$ correspond to electric $Ej$ multipole transitions, and those with $\Lambda=0$ correspond to magnetic $Mj$ multipole transitions. The terms with $\Lambda=-1$ do not contribute to multipole transitions, because $\mathbf{f}_\lambda(\mathbf{k})\mathbf{Y}_{JM}^{(-1)}(\theta_{k},\phi_k)=0$. Let us write an explicit expression for this scalar product separating the dependence on $ \phi_k $. The polarization vector is defined in formula \eqref{VecPol}. Then we have
\begin{equation}
	\bigg(\mathbf{f}_\lambda(\mathbf{k}) \mathbf{Y}_{JM}^{(\Lambda)*}(\theta_{k},\phi_{k})\bigg) = e^{-iM\phi_k}\Theta_{JM;\lambda}^{(\Lambda)}(\theta_{k}).
\end{equation}
The following notation has been introduced
\begin{equation}\label{Theta}
	\begin{split}
		\Theta_{JM;\lambda}^{(\Lambda)}(\theta_{k}) &= N P^M_{J+\La}(\cos\theta_{k})\bigg(\sqrt{\dfrac{J}{2J+1}}\bigg)^{\La}\big[\dfrac{n_3-\la}{2} C_{J+\La M+1 1 -1}^{J M} - \dfrac{n_3+\la}{2}C_{J+\La M-1 1 1}^{J M} - \dfrac{n_\perp}{\sqrt{2}}C_{J+\La M 1 0}^{J M}\big] + \\ 
		&+ \La N P^M_{J-\La}(\cos\theta_{k})\sqrt{\dfrac{J+1}{2J+1}} \big[\dfrac{n_3-\la}{2} C_{J-\La M+1 1 -1}^{J M} - \dfrac{n_3+\la}{2}C_{J-\La M-1 1 1}^{J M} - \dfrac{n_\perp}{\sqrt{2}}C_{J-\La M 1 0}^{J M}\big] ,\\
		Y_{JM}(\theta_k, \phi_k) &:=N P^M_{J}(\cos\theta_{k}) e^{i M \phi_k},
	\end{split}
\end{equation}
where $ N $ is the normalization factor of the spherical harmonics, and $ P^M_{J}(x)$ are associated Legendre polynomials.

Note that the quarkonium wave function is concentrated in the region $ r < r_B $. Then, if the transition energy is such that $ k_0r_B \ll 1 $, the following asymptotics can be employed  
\begin{equation}
	j_J(\la r) \approx \dfrac{J!}{(2J+1)!}(\la r)^J.
\end{equation}
Therefore the radial integrals can be replaced by the approximate expressions
\begin{equation}
	\begin{split}
		R_{n'l'nl}^J \approx \dfrac{J!}{(2J+1)!} \bigg(\frac{k_0}{2} \bigg) ^J r^{J-1}_{n'l'nl}, \qquad	r^J_{n'l'nl} := \int_0^{\infty} dr  r^{J + 2} R_{n'l'}^*(r) R_{nl}(r).
	\end{split}
\end{equation}

To simplify the integrals in $B_{J}^{(1)}$, we rewrite the Wronskian using the radial Schr{\"o}dinger equation as
\begin{equation}
	(r^2 W)' = r^2 R_{nl}R^*_{n'l'}\Big[ \frac{l(l+1) - l'(l'+1)}{ r^2} + m_q (E_{n'l'm'}-E_{nlm})\Big].
\end{equation}
Then in the leading order with respect to $k_0 r_B$ we derive
\begin{equation}
	B_{J}^{(1)}  =  -i (E_{n'l'm'}-E_{nlm})((-1)^{J} - 1) \frac{J!}{2(2J+1)!}\sqrt{\frac{(J+1)(2J+1)(2l+1)}{4 \pi J(2l'+1)}} \big(\frac{k_0}{2}\big)^{J-1} r^J_{n'l'nl}.
\end{equation}
This matrix element describes the $EJ$ transitions in the long-wave approximation.

\section{Matrix element $K_\alpha^-$}\label{BigKappaEvaluation}
Now we calculate the matrix element 
\begin{equation}
	K_\ga  ^{-}= \frac{1}{2m_q}\langle\text{out}| \hat{c}_\ga   \big(\Psi_{\ga   i}(\mathbf{R}-\frac{1}{2}\mathbf{r}) -\Psi_{\ga   i}(\mathbf{R}+\frac{1}{2}\mathbf{r}) \big)P_i |\text{in}\rangle.
\end{equation}
Its evaluation is reduced to the evaluation of the matrix element
\begin{equation}
\begin{split}
	\langle\text{out}| \hat{c}_\ga \Psi_{\ga i}(\mathbf{R}-b\mathbf{r}) \hat{P}_i |\text{in}\rangle=\frac{\chi_\ga e^{i\lambda \phi_k}}{\sqrt{2k_0}}\int\frac{ d\mathbf{P}}{(2\pi)^{3/2}} g(\mathbf{P})\langle \mathbf{P}'|e^{i\mathbf{R}\mathbf{k}}\hat{P}_i|\mathbf{P}\rangle \langle n'l'm'|f_{\lambda i}(\mathbf{k})e^{-ib\mathbf{r}\mathbf{k}} |nlm\rangle.
\end{split}
\end{equation}
The main technical complexity comes from the calculation of the matrix element
\begin{equation}\label{K_inner_matr_element}
\begin{split}
	\langle n'l'm'|\mathbf{f}_{\lambda}(\mathbf{k})e^{-ib\mathbf{r}\mathbf{k}}|nlm\rangle=\int  r^2 dr \sin\theta d\theta d\vf  R_{n'l'}^*(r)Y_{l'm'}^*(\theta,\vf)R_{nl}(r)Y_{lm}(\theta,\vf)\mathbf{f}_\lambda (\mathbf{k})e^{-ib\mathbf{k}\mathbf{r}}.
\end{split}
\end{equation}
This matrix element is calculated along the lines of the procedure presented in the previous section. However, in this case it is more convenient to use a different form of the mode functions
\begin{equation}\label{multipoles}
	\bs\psi^\La_{JM}(k_0,\spx)=
	\left\{
	\begin{array}{ll}
		j_J(k_0r)\mathbf{Y}^J_{JM}(\theta,\vf), & \hbox{$\La=0$,} \\
		-\sqrt{\frac{J}{2J+1}} j_{J+1}(k_0r)\mathbf{Y}^{J+1}_{JM}(\theta,\vf) +\sqrt{\frac{J+1}{2J+1}}
		j_{J-1}(k_0r)\mathbf{Y}^{J-1}_{JM}(\theta,\vf), & \hbox{$\La=1$}.
	\end{array}
	\right.
\end{equation}
We substitute the multipole expansion \eqref{multipole_expension} with the mode functions \eqref{multipoles} into \eqref{K_inner_matr_element}. In this procedure the following angular integrals arise
\begin{equation}
	\begin{split}
		&\int  \sin\theta d\theta d\vf Y_{l'm'}^*(\theta,\vf)Y_{lm}(\theta,\vf) \mathbf{Y}_{JM}^{L}(\theta,\phi) = \sqrt{\frac{(2L+1)(2l+1)}{4\pi(2l'+1)}} \sum_{\s} C_{l0 L0}^{l'0} C_{lm L M-\s}^{l'm'} C_{1\s L M-\s}^{J M} \spe_\s =\\
		& = (-1)^{l'+L}\sqrt{\frac{(2l+1)}{4\pi}} \sum_{\s} C_{l'0 l0}^{L0} C_{lm L M-\s}^{l'm'} C_{1\s L M-\s}^{J M} \spe_\s =: \sum_{\s=\pm 1, 0} \Phi_{ l'l,JM}^{L,\s} C_{lm L M-\s}^{l'm'} \spe_\s .
	\end{split}
\end{equation}
Thus
\begin{equation}
\begin{split}
	\langle n'l'm'|&\mathbf{f}_{\lambda}(\mathbf{k})e^{-i\frac{\xi}{2}\mathbf{r}\mathbf{k}}|nlm\rangle =	4\pi \sum_{J\Lambda M}(-\xi)^{J+\Lambda} i^{J+\Lambda}\bigg(\mathbf{f}_\lambda(\mathbf{k}) \mathbf{Y}_{JM}^{(\Lambda)*}(\theta_{k},\phi_{k})\bigg)\times
	\\
	\times\bigg[& \bigg(\sqrt{\frac{J}{2J+1}}\bigg)^{\La} \tilde{R}_{n'l'nl}^{J+\La}  \sum_{\s=\pm 1, 0} \Phi_{ l'l,JM}^{J+\La,\s} C_{lm J+\La M-\s}^{l'm'} \spe_\s - \Lambda\sqrt{\frac{J+1}{2J+1}} \tilde{R}_{n'l'nl}^{J-\La}  \sum_{\s=\pm 1, 0} \Phi_{ l'l,JM}^{J-\La,\s} C_{lm J-\La M-\s}^{l'm'} \spe_\s
	\bigg],
\end{split}
\end{equation}
where the following notation has been introduced
\begin{equation}
	 \tilde{R}_{n'l'nl}^{J}  = \int_0^\infty r^2 dr j_{J}\bigg(\frac{k_0r}{2}\bigg)  R_{n'l'}^*(r) R_{nl}(r).
\end{equation}
So, eventually we arrive at
\begin{equation}\label{MultipoleExpansionForBigKappa}
\begin{split}
	&K_\alpha^{-}=M_q^{-1}\langle\text{out}| \hat{c}_\alpha \big[\Psi_{\alpha i}(\mathbf{R}-\frac{m_q}{M_q}\mathbf{r}) -\Psi_{\alpha i}(\mathbf{R}+\frac{m_q}{M_q}\mathbf{r}) \big]P_i |\text{in}\rangle
	=\\=
	&\frac{4\pi \chi_\ga e^{i\lambda\phi_k}}{\sqrt{2k_0}}\int\frac{ d\mathbf{P}}{(2\pi)^{3/2}} g(\mathbf{P}) \langle \mathbf{P}'|e^{i\mathbf{R}\mathbf{k}}|\mathbf{P}\rangle  \sum_{J\Lambda M}i^{J+\Lambda} \bigg(\mathbf{f}_\lambda(\mathbf{k}) \mathbf{Y}_{JM}^{(\Lambda)*}(\theta_{k},\phi_{k})\bigg)	\bigg(\mathbf{W}^{(\Lambda)}_{JM}\mathbf{P}\bigg),
\end{split}
\end{equation}
where
\begin{equation}\label{W_JM}
\begin{split}
	\mathbf{W}^{(\Lambda)}_{JM } = \frac{(-1)^{J+\Lambda}-1}{2 m_q} &\bigg[\bigg(\sqrt{\frac{J}{2J+1}}\bigg)^{\La} \tilde{R}_{n'l'nl}^{J + \La}  \sum_{\s=\pm 1, 0} \Phi_{ l'l,JM}^{J+\La,\s} C_{lm J+\La M-\s}^{l'm'} \spe_\s - \\
	& - \Lambda\sqrt{\frac{J+1}{2J+1}}\tilde{R}_{n'l'nl}^{J-\La}  \sum_{\s=\pm 1, 0} \Phi_{ l'l,JM}^{J-\La,\s} C_{lm J-\La M-\s}^{l'm'} \spe_\s
\bigg].
\end{split}
\end{equation}
\normalsize
The selection rules \eqref{WSelRul} are evident from this expression.

The following notation is also useful:
\begin{equation}\label{W_JMFinal}
	\begin{split}
		|W^{(0)}_{JM}|^2 &= \frac{1-(-1)^J}{2m^2_q} | \tilde{R}_{n'l'nl}^{J} |^2 \sum_{\s=\pm 1, 0}|\Phi_{ l'l,JM}^{J,\s}|^2,  \\
		|W^{(1)}_{JM}|^2 &= \frac{1-(-1)^{J+1}}{2m^2_q} \bigg[\frac{J}{2J+3}|\tilde{R}_{n'l'nl}^{J+1} |^2 \sum_{\s=\pm 1, 0}|\Phi_{ l'l,JM}^{J+1,\s}|^2 + \frac{J+1}{2J-1}|\tilde{R}_{n'l'nl}^{J-1}|^2 \sum_{\s=\pm 1, 0}|\Phi_{ l'l,JM}^{J-1,\s}|^2\bigg].
	\end{split}
\end{equation}
These expressions are employed in \eqref{PrTwSumm} and \eqref{PrTwSummSiml}.

If the condition $k_0r_B \ll 1$ is satisfied, the radial integral can be simplified as
\begin{equation}
	 \tilde{R}_{n'l'nl}^{J}  \approx \dfrac{J!}{(2J+1)!}\big(\frac{k_0}{2}\big)^{J} r_{n'l'nl}^{J}.
\end{equation}
Then the leading contribution to the probability \eqref{PrTwSumm} or \eqref{PrTwSummSiml} in the long-wave approximation coming from the center-of-mass dynamics corresponds to $J = |l'-l|$ for the term $|W^{(0)}_{JM}|^2$, $J = |l'-l| - 1$ for the first term in $|W^{(1)}_{JM}|^2$, and $J = |l'-l| + 1$ for the second one. Thus, one can see, that there are three terms in \eqref{W_JMFinal} of the same order in $k_0 r_B $.


\section{Integrals over $\phi_{k}$}\label{phiint}
We need to calculate integrals of the form 
\begin{equation}\label{AngInt}
	\int_0^{2\pi}  d\phi_{k_1} d\phi_{k_2} e^{- \frac{\De_{12}\spk_\perp^2}{8\s^2_c}} e^{i\mathbf{b_\perp}\De_{12}\spk_\perp} e^{im_\gamma (\phi_{k_1} - \phi_{k_2})} e^{-i M_1 \phi_{k_1}}e^{i M_2 \phi_{k_2}}.
\end{equation}
Taking into account that
\begin{equation}
	\begin{split}
		\mathbf{b_\perp}\De_{12}\spk_\perp = k_\perp^0 b_\perp[\cos(\phi_{k_1} -\phi_b) - \cos(\phi_{k_2} - \phi_b)], \\
		\De_{12}\spk_\perp^2 = 2( k_\perp^0)^2 - 2 ( k_\perp^0)^2\cos(\phi_{k_1}- \phi_{k_2}),
	\end{split}
\end{equation}
and Jacobi-Anger identity
\begin{equation}
	e^{z\cos\vf}=\sum_{n=-\infty}^{\infty} e^{in\vf} I_n(z),
\end{equation}
we rewrite the integral \eqref{AngInt} as
\begin{equation}
	e^{-\frac{( k_\perp^0)^2}{4\s_c^2}}\sum_{n=-\infty}^{\infty} I_n\big(\frac{( k_\perp^0)^2}{4\s_c^2}\big) \int_0^{2\pi}    d\phi_{k_1} d\phi_{k_2} e^{i[k_\perp^0b_\perp\cos(\phi_{k_1} -\phi_b) -(M_1-m_\ga-n)\phi_{k_1}]} e^{-i[k_\perp^0b_\perp\cos(\phi_{k_2} -\phi_b) -(M_2-m_\ga-n)\phi_{k_2}]}.
\end{equation}
Employing the integral representation for the Bessel function 
\begin{equation}
	J_m(z)=i^{-m}\int_{0}^{2\pi}\frac{d\vf}{2\pi} e^{i(z\cos\vf -m\vf)},
\end{equation}
we obtain 
\begin{equation}
	(2\pi)^2 e^{-\frac{( k_\perp^0)^2}{4\s_c^2}} \sum_{n=-\infty}^{\infty} I_n\big(\frac{( k_\perp^0)^2}{4\s_c^2}\big)i^{M_1-M_2}e^{i\phi_b(M_2-M_1)}J_{M_1-m_\ga-n}(k_\perp^0b_\perp)J_{M_2-m_\ga-n}(k_\perp^0b_\perp).
\end{equation}

The integrals over $\phi_{k_{1,2}}$ for terms standing on the last line in \eqref{PrTw1} can also be evaluated. For this purpose, let us introduce the notation
\begin{equation}
\begin{split}
	W_\perp^1(\cos\phi_{W_1},\sin\phi_{W_1},0):=&(W^{(\Lambda_1)}_{1,J_1M_1},W^{(\Lambda_1)}_{2,J_1M_1},0),\quad W_\perp^2(\cos\phi_{W_2},\sin\phi_{W_2},0):=(W^{*(\Lambda_2)}_{1,J_2M_2},W^{*(\Lambda_2)}_{2,J_2M_2},0),\\
	\mathbf{T} := T_\perp(\cos\phi_{T},\sin\phi_{T},0):=& \big(C_{lm J_1M_1}^{l'm'} B_{J_1}^{(\La_1)}W^{*(\La_2)}_{1,J_2M_2} - C_{lm J_2M_2}^{l'm'}B_{J_2}^{*(\La_2)} W^{(\La_1)}_{1,J_1M_1}, C_{lm J_1M_1}^{l'm'} B_{J_1}^{(\La_1)}W^{*(\La_2)}_{2,J_2M_2} - \\
	&- C_{lm J_2M_2}^{l'm'}B_{J_2}^{*(\La_2)} W^{(\La_1)}_{2,J_1M_1}, 0 \big).
\end{split}
\end{equation}
Then, for the terms standing in the last line of equation \eqref{PrTw1}, we have
\begin{equation}
\begin{split}
	(\De_{12}\spk_\perp \cdot \mathbf{W}^{(\La_1)}_{J_1M_1}) (\De_{12}\spk_\perp\cdot\mathbf{W}^{*(\La_2)}_{J_2M_2})&=(k_\perp^0)^2 W_\perp^1 W_\perp^2(\cos(\phi_1-\phi_{W_1})-\cos(\phi_2-\phi_{W_1})) \times \\
	&\times (\cos(\phi_1-\phi_{W_2})-\cos(\phi_2-\phi_{W_2})),\\
	\frac{\De_{12}\spk_\perp}{2}\mathbf{T}&=\frac{1}{2}k_\perp^0 T_\perp(\cos(\phi_1-\phi_{T})-\cos(\phi_2-\phi_{T})).
\end{split}
\end{equation}
Now, employing similar technique to the one used to derive \eqref{PrTw2}, we obtain
\begin{equation}
\begin{split}
	& P^{(Tw)}_{n'l'm',nlm} ={\frac{8\pi^2 \e q^2 \s_\perp^2}{\s_3 \tilde{k}^2_{3}(k_\perp^0)}} e^{-\frac{( k_\perp^0)^2}{4\s_c^2}} e^{-\frac{( \tilde{k}_{3}(k_\perp^0)-k_3^0)^2}{2\sigma_3^2}}\sum_{J_1 J_2 } \sum_{M_1 M_2} \sum_{\La_1\La_2} \sum_{n=-\infty}^{\infty} i^{J_1+\La_1+M_1-J_2-\La_2-M_2} e^{-i(M_1-M_2)\phi_{b}}\times \\
	&\times \Theta_{J_1M_1;\la_0}^{(\La_1)}(\theta_{k}^0) \Theta_{J_2M_2;\la_0}^{*(\La_2)}(\theta_{k}^0)  
	I_n\big(\frac{( k_\perp^0)^2}{4\s_c^2}\big)\Big[ J_{(M_1-m_\ga-n)} J_{(M_2-m_\ga-n)}[C_{lm J_1M_1}^{l'm'} C_{lm J_2M_2}^{l'm'}B_{J_1}^{(\La_1)}B_{J_2}^{*(\La_2)} +\\
	& + {\s_c^2} \mathbf{W}^{(\La_1)}_{J_1M_1}\mathbf{W}^{*(\La_2)}_{J_2M_2} -\frac{1}{4}(k_\perp^0)^2 W_\perp^1 W_\perp^2   
	\cos(\phi_{W_1}-\phi_{W_2}) ]+ \frac{1}{8}(k_\perp^0)^2 W_\perp^1 W_\perp^2  \big( \frac{1}{2}e^{-i(\phi_{W_1}+\phi_{W_2}-2\phi_b)} \times\\
	& \times ( J_{M_2-m_\ga-n} J_{M_1-m_\ga-n-2} +  J_{M_1-m_\ga-n} J_{M_2-m_\ga-n-2} )
	+ \frac{1}{2}e^{i(\phi_{W_1}+\phi_{W_2}-2\phi_b)} (J_{M_2-m_\ga-n}J_{M_1-m_\ga-n+2} +\\
	& +  J_{M_1-m_\ga-n} J_{M_2-m_\ga-n+2}) +(e^{i(\phi_b-\phi_{W_1})}J_{M_1-m_\ga-n-1}-e^{-i(\phi_b-\phi_{W_1})} J_{M_1-m_\ga-n+1}) \times \\
	& \times (e^{-i(\phi_b-\phi_{W_2})}J_{M_2-m_\ga-n-1} - e^{i(\phi_b-\phi_{W_2})}  J_{M_2-m_\ga-n+1}) +(e^{i(\phi_b-\phi_{W_2})}J_{M_1-m_\ga-n-1}-\\
	& - e^{-i(\phi_b-\phi_{W_2})} J_{M_1-m_\ga-n+1})(e^{-i(\phi_b-\phi_{W_1})}J_{M_2-m_\ga-n-1} - e^{i(\phi_b-\phi_{W_1})}  J_{M_2-m_\ga-n+1})
	\big)+\\
	&+\frac{i}{4}k_\perp^0 T_\perp
	( e^{-i(\phi_b-\phi_{T})} J_{M_2-m_\ga-n}J_{M_1-m_\ga-n+1} - e^{i(\phi_b-\phi_{T})} J_{M_2-m_\ga-n} J_{M_1-m_\ga-n-1} +\\
	&+e^{i(\phi_b-\phi_{T})} J_{M_1-m_\ga-n} J_{M_2-m_\ga-n+1} - e^{-i(\phi_b-\phi_{T})} J_{M_1-m_\ga-n} J_{M_2-m_\ga-n-1}  )  \Big].
\end{split}
\end{equation}

\begin{center}
	
\end{center}

\end{document}